\documentclass[a4paper,11pt]{article}
\usepackage{jheppub} 
\usepackage[T1]{fontenc} 
\usepackage{multirow}
\usepackage{color}
\usepackage{caption}
\usepackage{rotating}
\usepackage{booktabs}
\usepackage{dcolumn}
\usepackage{array}
\usepackage{graphicx}
\usepackage{multirow}
\usepackage{lineno}
\usepackage{siunitx}
\usepackage{amsmath}
\usepackage{bm}
\usepackage{mathrsfs}
\usepackage{overpic}
\usepackage{indentfirst}

\newcommand{\BESIIIorcid}[1]{\href{https://orcid.org/#1}{\hspace*{0.1em}\raisebox{-0.45ex}{\includegraphics[width=1em]{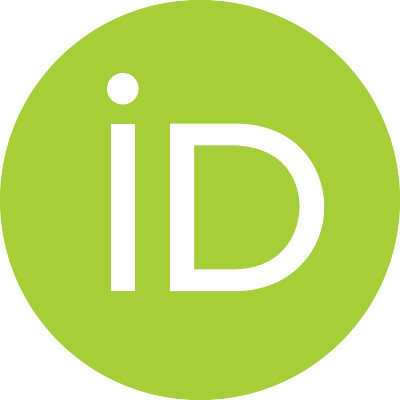}}}}

\title{\protect\boldmath Measurement of the Born cross section for $\boldsymbol{e^+e^- \to p K^- K^- \bar{\Xi}^{+}}$ at $\boldsymbol{\sqrt{s} = 3.5 - 4.9}$ GeV}
\collaborationImg{\includegraphics[width=0.25\textwidth]{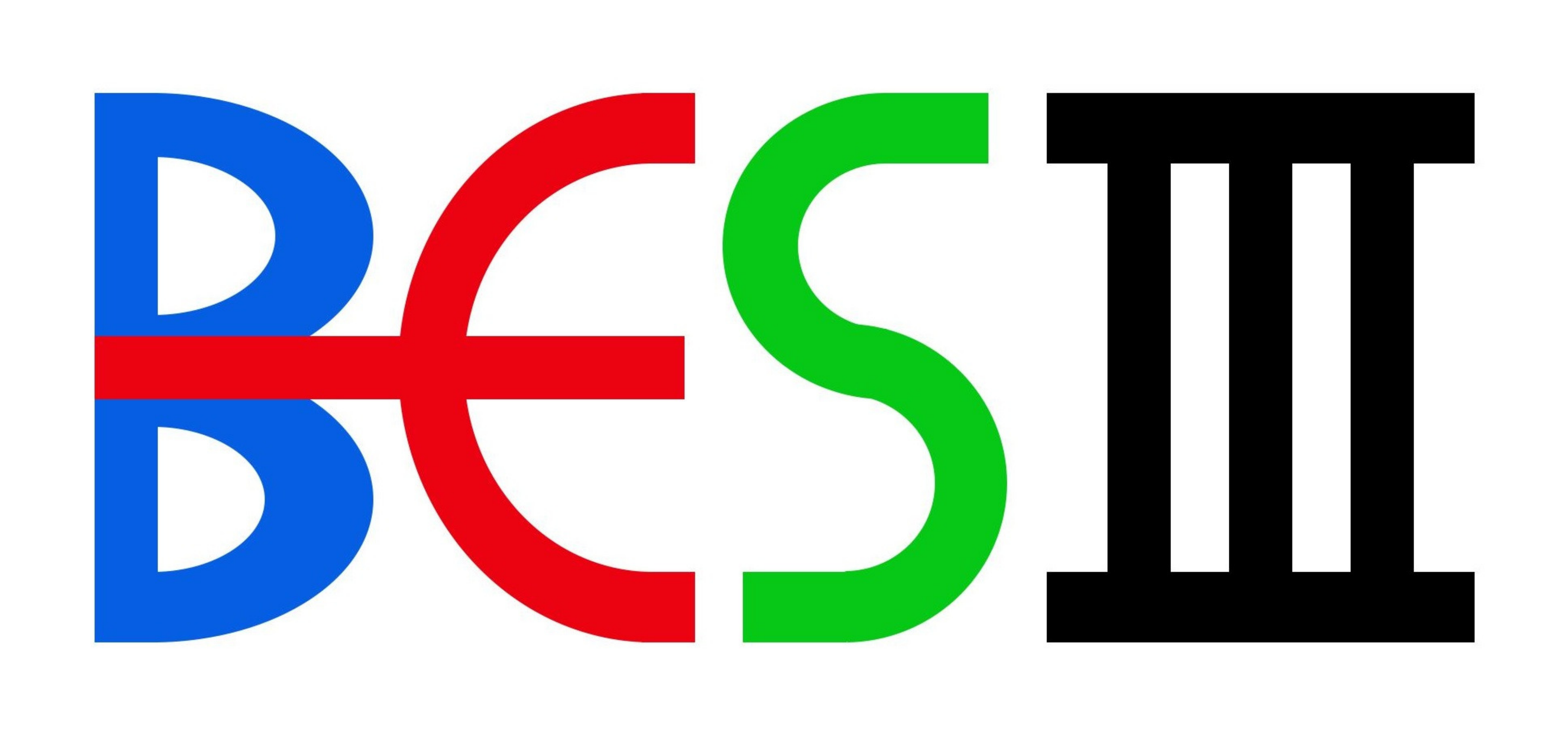}}
\collaboration{The BESIII collaboration}
\emailAdd{besiii-publications@ihep.ac.cn}

\begin{document}

\abstract{Using $e^+ e^-$ collision data corresponding to a total integrated luminosity of 20 ${\rm fb}^{-1}$ collected with the BESIII detector at the BEPCII collider,
we present a measurement of the Born cross section for the process $e^+e^- \to p K^-K^-\bar{\Xi}^{+}$ at 39 center-of-mass energies between 3.5 and 4.9 GeV with a partial reconstruction technique.
By performing a fit to the dressed cross section of $e^{+}e^{-}\to p K^- K^-\bar{\Xi}^{+}$ with a power law function for continuum production and one resonance at a time for the $\psi(3770)$, $\psi(4040)$, $\psi(4160)$, $\psi(4230)$, $\psi(4360)$, $\psi(4415)$ or $\psi(4660)$, respectively,
the upper limits for the product of partial electronic width and branching fraction into the final state $p K^- K^- \bar{\Xi}^+$ for these resonances are determined at the $90\%$ confidence level.
}

\maketitle
\flushbottom

\section{Introduction}
\noindent
The study of baryon decays of vector charmonium(-like) resonances provides a test of quantum chromodynamics (QCD)~\cite{ccbar1,ccbar2}.
Below the open charm threshold, the mass spectrum of the conventional charmonium resonances is consistent with the predictions from the potential quark model~\cite{qm}.
Above the open charm threshold, the model predicts six vector charmonium states from the threshold to 4.9 GeV, ordered as the $1D$, $3S$, $2D$, $4S$, $3D$, and $5S$ states, while far more vector states have been observed in this energy region.
Several states, such as  the $\psi(4040)$, $\psi(4160)$, and $\psi(4415)$, are seen as clear peaks in the center-of-mass energy-dependent inclusive hadronic cross section and predominantly decay to open charm final states~\cite{besi}.
Other states, such as the $\psi(4230)$, $\psi(4360)$, and $\psi(4660)$, are mainly observed in decay channels with hidden-charm intermediate states and are produced via initial state radiation (ISR) processes at BaBar and Belle~\cite{belle1,belle2,belle3,belle4,belle5,belle6,belle7,babar1,babar2,Wang:2025dur} or via direct production in $e^+e^-$ annihilation at CLEO~\cite{cleo} and BESIII~\cite{bes1,bes2}.
These vector states might not be resonances with simple $c\bar{c}$ quark content, and many hypotheses, such as hybrid, multiple-quark state and molecular models, have been proposed to interpret them~\cite{Close:2005iz,Xia:2015mga,Chen:2016qju,Qian:2021neg,Bai:2023dhc,Yan:2023yff,Dai:2023vsw}.
No solid conclusions can be made at present and the nature of these vector charmonium(-like) states remains uncertain~\cite{Yuan:2021wpg}.
This reflects our limited knowledge of the strong interaction in the non-perturbative region.
To solve these puzzles, more and improved experimental measurements are needed.
Among these measurements, charmonium states decaying into hyperon-antihyperon pairs, dominated by three-gluon or one-photon processes, are promising due to the simple topologies of their final states compared to three-meson production.
However, studies of substantial correlations between these charmonium(-like) states and the production of baryonic final states above open charm threshold are still insufficient.
Although many experimental studies of  baryonic final states have been performed by the BESIII and Belle collaborations\cite{
Ablikim:2013pgf, BESIII:2016ssr,BESIII:2016nix,Wang:2018kdh,BESIII:2019dve,Ablikim:2019kkp,BESIII:2020ktn,BESIII:2021aer,BESIII:2021ccp, BESIII:2021gca,
BESIII:2021cvv,Wang:2022zyc,BESIII:2022mfx,BESIII:2022lsz,BESIII:2022kzc, BESIII:2023lkg,Liu:2023xhg,BESIII:2023euh,BESIII:2023rwv,BESIII:2023rse,BESIII:2024ogz,BESIII:2024dmr,BESIII:2024ues,BESIII:2024umc,BESIII:2024gql,BESIII:2025yzk,Zhang:2025qmo}, the only established decay is $\psi(4660) \to \Lambda^+_c \bar{\Lambda}^-_c$~\cite{belle5}. While there is also evidence for $\psi(3770) \to \Lambda \bar{\Lambda}$~\cite{BESIII:2021ccp}, $\psi(3770) \to \Xi^-\bar{\Xi}^+$~\cite{BESIII:2023rse}, and the three-body baryonic decay $\psi(4160) \to K^- [\Lambda/\Sigma^0] \bar{\Xi}^+$~\cite{BESIII:2024ogz} with an extra pseudoscalar meson, our knowledge is still lacking.
Thus, more precise measurements of exclusive cross sections of $e^+e^-$ to baryonic final states above the open charm threshold could provide additional insight into the nature of these vector charmonium(-like) states.

In this paper, we present a measurement of the Born cross sections for $e^+e^-\to p K^-K^-\bar{\Xi}^{+}$ using data sets at center-of-mass (CM) energies $\sqrt{s}$ between 3.5 GeV and 4.9 GeV corresponding to a total integrated luminosity of 20 ${\rm fb}^{-1}$~\cite{lumi1,lumi2} collected by the BESIII detector~\cite{besiii1, besiii2} at the BEPCII collider~\cite{Yu:IPAC2016-TUYA01}.
Compared to previous studies, this process is a four-body interaction that includes two pseudoscalar mesons.
Throughout this paper, the charge conjugated channel is always included.
The possible resonances are studied by fitting the lineshape of the dressed cross section of the process $e^+e^-\to p K^-K^-\bar{\Xi}^{+}$.
The products of branching fraction and electronic partial width for
possible charmonium(-like) states, i.e. $\psi(3770)$, $\psi(4040)$, $\psi(4160)$, $\psi(4230)$, $\psi(4360)$ and $\psi(4660)$ decaying to $pK^-K^-\bar{\Xi}^{+}$, as well as their upper limits at the 90\% confidence level (C.L.) are provided.

\section{BESIII detector and Monte Carlo simulation}
\noindent
The BESIII detector records symmetric $e^+e^-$ collisions provided by the BEPCII storage ring in the CM energy range from 1.84 GeV to 4.95 GeV, with a peak luminosity of $1.1 \times 10^{33}\;\text{cm}^{-2}\text{s}^{-1}$ achieved at $\sqrt{s} = 3.773$ GeV.
BESIII has collected large data samples in this energy region.
The cylindrical core of the BESIII detector covers 93\% of the full solid angle and consists of a helium-based multilayer drift chamber (MDC), a time-of-flight system (TOF), and a CsI(Tl) electromagnetic calorimeter~(EMC), which are all enclosed in a superconducting solenoidal magnet providing a 1.0~T magnetic field.
The solenoid is supported by an octagonal flux-return yoke with resistive plate counter muon identification modules interleaved with steel.
The charged-particle momentum resolution at $1~{\rm GeV}/c$ is $0.5\%$, and the ${\rm d}E/{\rm d}x$ resolution is $6\%$ for electrons from Bhabha scattering.
The EMC measures photon energies with a resolution of $2.5\%$ ($5\%$) at $1$~GeV in the barrel (end cap) region.
The time resolution in the plastic scintillator TOF barrel region was 68 ps, while that in the end cap region was 110 ps. The end cap TOF
system was upgraded in 2015 using multigap resistive plate chamber
technology providing a time resolution of 60~ps, which benefits 64\% of the data used in this analysis~\cite{etof1,etof2,etof3}.

Monte Carlo (MC) simulated data samples produced with a \textsc{geant4}-based~\cite{geant41, geant42} package, which includes the geometric description of the BESIII detector and the detector response, are used to determine detection efficiencies and to estimate backgrounds. The simulation models the beam energy spread and ISR in the $e^+e^-$ annihilations with the generator \textsc{kkmc}~\cite{kkmc1, kkmc2}.
The inclusive MC sample includes the production of open charm processes, the ISR
production of vector charmonium(-like) states, and the continuum processes incorporated in \textsc{kkmc}.
All particle decays are modeled with \textsc{evtgen}~\cite{evtgen1, evtgen2} using branching fractions either taken from the Particle Data Group (PDG)~\cite{PDG2022}, when available, or otherwise estimated with \textsc{lundcharm}~\cite{lundcharm}. Final state radiation~(FSR)
from charged particles is incorporated using the \textsc{photos} package~\cite{photos2}.
To determine the detection efficiencies and evaluate the ISR factors in the $e^+e^- \to p K^-K^-\bar{\Xi}^+$, $\bar{\Xi}^+ \to$ anything process, exclusive MC samples with $6 \times 10^5$ signal events are generated for each of the 39 CM energies from 3.5 GeV to 4.9 GeV using a phase space (PHSP) model and incorporating ISR effects.

\section{Event selection}
\noindent
A partial reconstruction method is used to extract the signal process, where only the proton and two kaons are reconstructed in each event and the presence of the $\bar{\Xi}^+$ is inferred through the missing mass.
The tracks from charged particles must at least fulfill the following criteria. The polar angle for each track in the MDC must satisfy $|\cos \theta| < 0.93$, where $\theta$ is the angle between the direction of the track and the positron beam direction.
To suppress backgrounds, decay vertices are limited in the distance $V_r < 1 {\rm ~cm}$ in the radial and $|V_z| < 10 {\rm ~cm}$ in the axial direction from the interaction point.
Particle identification (PID) probabilities, Prob$(H)$, $H = p,~K,~\pi$, are determined from the ionization energy loss measured in the MDC combined with the time of flight measured with TOF.
Tracks are identified as protons if the proton hypothesis has the greatest probability (Prob$(p)$ > Prob$(K)$ and Prob$(p)$ > Prob$(\pi)$), while charged kaons are identified by requiring Prob$(K)$ > Prob$(p)$ and Prob$(K)$ > Prob$(\pi)$.
Events with exactly one $p$ candidate and two $K^-$ candidates, denoted in the following as $K_1$ and $K_2$, are kept for further analysis.
If an extra candidate for the identified tracks is present, the event is discarded.

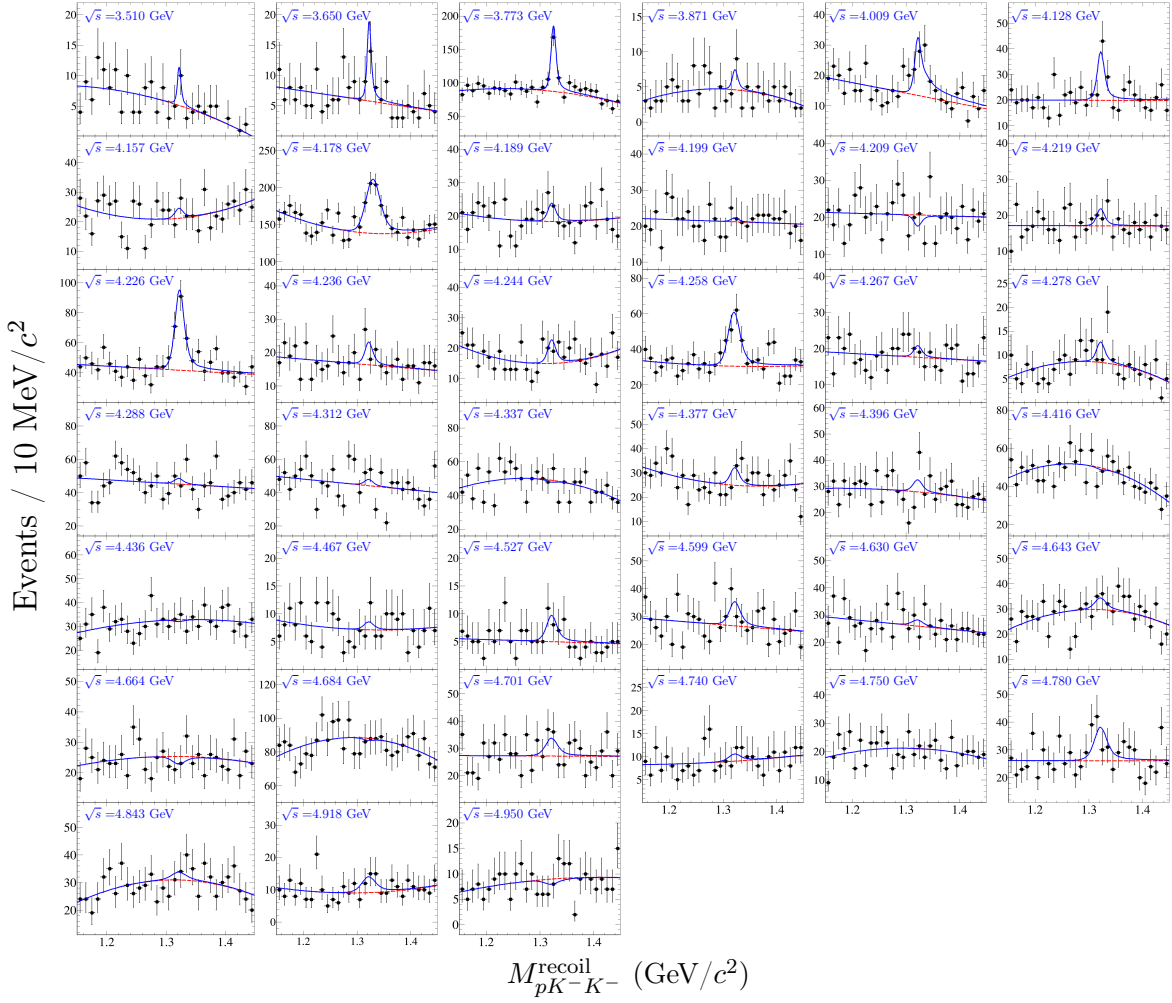
\begin{figure}[!htp]
    \centering
    \begin{overpic}[scale=0.78]{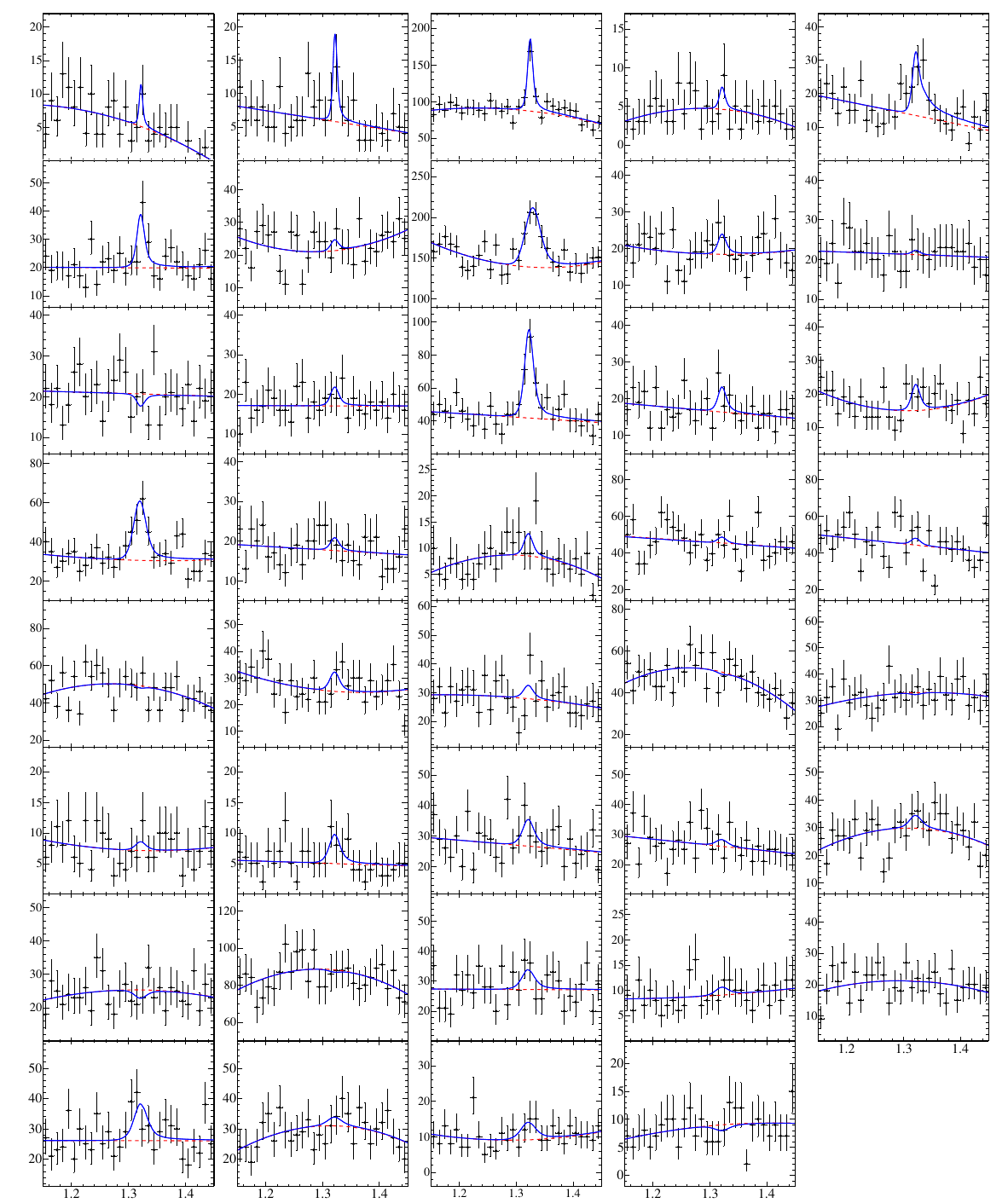}
    \put( 4.4,97.5){\color{blue}\scalebox{0.6}{$\sqrt{s}=$3.510 GeV}}
    \put(20.5,97.5){\color{blue}\scalebox{0.6}{$\sqrt{s}=$3.650 GeV}}
    \put(36.6,97.5){\color{blue}\scalebox{0.6}{$\sqrt{s}=$3.773 GeV}}
    \put(52.7,97.5){\color{blue}\scalebox{0.6}{$\sqrt{s}=$3.871 GeV}}
    \put(68.8,97.5){\color{blue}\scalebox{0.6}{$\sqrt{s}=$4.009 GeV}}
    \put( 4.4,85.3){\color{blue}\scalebox{0.6}{$\sqrt{s}=$4.128 GeV}}
    \put(20.5,85.3){\color{blue}\scalebox{0.6}{$\sqrt{s}=$4.157 GeV}}
    \put(36.6,85.3){\color{blue}\scalebox{0.6}{$\sqrt{s}=$4.178 GeV}}
    \put(52.7,85.3){\color{blue}\scalebox{0.6}{$\sqrt{s}=$4.189 GeV}}
    \put(68.8,85.3){\color{blue}\scalebox{0.6}{$\sqrt{s}=$4.199 GeV}}
    \put( 4.4,73.1){\color{blue}\scalebox{0.6}{$\sqrt{s}=$4.209 GeV}}
    \put(20.5,73.1){\color{blue}\scalebox{0.6}{$\sqrt{s}=$4.219 GeV}}
    \put(36.6,73.1){\color{blue}\scalebox{0.6}{$\sqrt{s}=$4.226 GeV}}
    \put(52.7,73.1){\color{blue}\scalebox{0.6}{$\sqrt{s}=$4.236 GeV}}
    \put(68.8,73.1){\color{blue}\scalebox{0.6}{$\sqrt{s}=$4.244 GeV}}
    \put( 4.4,60.9){\color{blue}\scalebox{0.6}{$\sqrt{s}=$4.258 GeV}}
    \put(20.5,60.9){\color{blue}\scalebox{0.6}{$\sqrt{s}=$4.267 GeV}}
    \put(36.6,60.9){\color{blue}\scalebox{0.6}{$\sqrt{s}=$4.278 GeV}}
    \put(52.7,60.9){\color{blue}\scalebox{0.6}{$\sqrt{s}=$4.288 GeV}}
    \put(68.8,60.9){\color{blue}\scalebox{0.6}{$\sqrt{s}=$4.312 GeV}}
    \put( 4.4,48.7){\color{blue}\scalebox{0.6}{$\sqrt{s}=$4.337 GeV}}
    \put(20.5,48.7){\color{blue}\scalebox{0.6}{$\sqrt{s}=$4.377 GeV}}
    \put(36.6,48.7){\color{blue}\scalebox{0.6}{$\sqrt{s}=$4.396 GeV}}
    \put(52.7,48.7){\color{blue}\scalebox{0.6}{$\sqrt{s}=$4.416 GeV}}
    \put(68.8,48.7){\color{blue}\scalebox{0.6}{$\sqrt{s}=$4.436 GeV}}
    \put( 4.4,36.5){\color{blue}\scalebox{0.6}{$\sqrt{s}=$4.467 GeV}}
    \put(20.5,36.5){\color{blue}\scalebox{0.6}{$\sqrt{s}=$4.527 GeV}}
    \put(36.6,36.5){\color{blue}\scalebox{0.6}{$\sqrt{s}=$4.599 GeV}}
    \put(52.7,36.5){\color{blue}\scalebox{0.6}{$\sqrt{s}=$4.630 GeV}}
    \put(68.8,36.5){\color{blue}\scalebox{0.6}{$\sqrt{s}=$4.643 GeV}}
    \put( 4.4,24.3){\color{blue}\scalebox{0.6}{$\sqrt{s}=$4.664 GeV}}
    \put(20.5,24.3){\color{blue}\scalebox{0.6}{$\sqrt{s}=$4.684 GeV}}
    \put(36.6,24.3){\color{blue}\scalebox{0.6}{$\sqrt{s}=$4.701 GeV}}
    \put(52.7,24.3){\color{blue}\scalebox{0.6}{$\sqrt{s}=$4.740 GeV}}
    \put(68.8,24.3){\color{blue}\scalebox{0.6}{$\sqrt{s}=$4.750 GeV}}
    \put( 4.4,12.1){\color{blue}\scalebox{0.6}{$\sqrt{s}=$4.780 GeV}}
    \put(20.5,12.1){\color{blue}\scalebox{0.6}{$\sqrt{s}=$4.843 GeV}}
    \put(36.6,12.1){\color{blue}\scalebox{0.6}{$\sqrt{s}=$4.918 GeV}}
    \put(52.7,12.1){\color{blue}\scalebox{0.6}{$\sqrt{s}=$4.950 GeV}}
    \put(-2,40.8){\rotatebox{90}{\scalebox{1.15}{Events / 10 MeV/$c^2$}}}
    \put(33.5,-3){\scalebox{1.15}{$M_{pK^-K^-}^{\rm recoil}$ (GeV/$c^2$)}}
    \end{overpic}
    \captionsetup{skip=25pt}
    \caption{Fits to the $M_{p K^- K^-}^{\rm recoil}$ spectra, where the black dots with error bars denote the data, the red dashed lines denote the background contribution, and the blue lines denote total fit curves.}
     \label{fit_pkk}
\end{figure}

To further suppress the background, a primary vertex fit for $p K^- K^-$ is performed, and events with $\chi^2 < 100$ are kept.
The number of anti-baryons $\bar{\Xi}^+$ at each energy point is determined using an unbinned maximum likelihood fit to the recoil mass of the $p K^- K^-$ system, defined as
\begin{equation}
\label{Recoil}
M^{\rm recoil}_{p K^- K^-} = \sqrt{\left[\sqrt{s} - \left(E_{p} + E_{K^-_1} + E_{K^-_2}\right)\right]^2 - \left| \vec{p}_{p} + \vec{p}_{K^-_1} + \vec{p}_{K^-_2}\right|^2},
\end{equation}
where $E_{p(K^-_1,K^-_2)}$ and $\vec{p}_{p(K^-_1, K^-_2)}$ are the energy and momentum of the proton (negative kaons), respectively.
The signal shape is modeled using a histogram of $M^{\rm recoil}_{p K^- K^-}$ from MC simulation, which is additionally convolved with a Gaussian function to account for different mass resolutions between data and MC simulation when the signal significance is greater than 3$\sigma$. 
Note that the resolution will be worse as the energy increases because the empirical beam energy spread $(0.9454 \sqrt{s} - 2.147)/\sqrt{2}$ MeV will dominate the effect.
Using a detailed background analysis in both data and MC simulation, the dominant background is determined to be from misidentified particles.  All backgrounds contribute either smooth shapes or shifted peaks away from the signal region in the recoil mass spectra, ensuring no contamination of the signal region. Consequently, the background shape is described by a first- or second-order Chebyshev polynomial function.
The fit results are presented in table~\ref{num_result} and shown in figure~\ref{fit_pkk}.
\section{Determination of Born cross section}
\noindent
The Born cross section ($\sigma^{\rm B}$) for $e^+e^- \to p K^-K^-\bar{\Xi}^+$ at each energy point is calculated as
\begin{equation}
\label{BCS}
  \sigma^{\rm B} = \frac{N_{\rm obs}}{2 \mathcal{L}\cdot\varepsilon \cdot(1 + \delta_{\rm ISR}) \cdot\frac{1}{\left|1 - \Pi\right|^2}},
\end{equation}
where $N_{\rm obs}$ is the signal yield, the factor of 1/2 averages the charge-conjugate channels, $\mathcal{L}$ is the integrated luminosity, $\varepsilon$ is the detection efficiency, $1 + \delta_{\rm ISR}$ is the ISR factor evaluated using a quantum electrodynamics calculation~\cite{qed}.  The procedure to determine both $\varepsilon$ and $1 + \delta_{\rm ISR}$ is iterated~\cite{isr} until the difference of their product between the last two rounds is less than $0.5\%$~\cite{BESIII:2023rse};  the iteration line shape is selected as the power-law function (described in section~\ref{sec66} ) only; and $\frac{1}{\left|1 - \Pi \right|^2}$ is the vacuum polarization factor estimated according to ref.~\cite{vp}.
The upper limit on the Born cross section is estimated at the $90\%$ C.L. Table~\ref{num_result} summarizes the numerical results.
\begin{table}[!htp]
    \centering
    \caption{Numerical results for the Born cross sections of $e^+ e^- \to p K^- K^- \bar{\Xi}^+$ at different energy points. Upper limits on $ N_{\rm obs}$ and $\sigma^{\rm B}$ at the $90\%$ C.L. are listed in parentheses if the significances are less than 3 $\sigma$. The first uncertainties are statistical and the second are systematic. The $S$ is the statistical significance of the signal.}
    \scalebox{0.75}{
    \begin{tabular}{c r@{$.$}l c c c r@{$~\pm~$}l r@{$~\pm~$}c@{$~\pm~$}l c}
    \hline
$\sqrt{s}$ (GeV)   &\multicolumn{2}{c}{$\mathcal{L}~({\rm pb}^{-1})$}    &$\varepsilon$ (\%)     &$1 + \delta_{\rm ISR}$    &$\frac{1}{\left| 1 - \Pi \right|^{2}}$     &\multicolumn{2}{c}{$N_{\rm obs}$}     &\multicolumn{3}{c}{$\sigma^{\rm B}$ (fb)}       &$S~(\sigma)$  \\
    \hline
3.510          &405&7          &15.27  &0.915  &1.045  &6.6&4.3~ ($<13.2$)        &$55.7$&$36.3$&$3.1~~(<111.4)$      &1.8 \\
3.650          &410&0          &25.61  &0.929  &1.021  &17.5&6.5                &$87.8$&$32.6$&$4.8$               &3.4 \\
3.773          &2931&8         &31.42  &0.968  &1.056  &157.1&23.4              &$83.5$&$12.4$&$4.6$               &7.2 \\
3.871          &219&2          &34.39  &0.959  &1.051  &5.0&4.7~ ($<12.5$)        &$32.9$&$30.9$&$1.8~~(<82.3)$       &1.2 \\
4.009          &481&96         &29.92  &1.030  &1.044  &63.8&16.1               &$205.7$&$51.9$&$11.3$             &4.6 \\
4.128          &401&5          &25.21  &1.077  &1.052  &43.8&13.0               &$191.0$&$56.7$&$10.5$             &3.9 \\
4.157          &408&7          &25.87  &1.082  &1.053  &8.0&10.2 ($<26.7$)       &$33.2 $&$48.1$&$1.8~~(<107.9)$     &$<1$ \\
4.178          &3194&5         &25.92  &1.095  &1.054  &291.2&81.3              &$152.4$&$42.5$&$8.4$              &6.4 \\
4.189          &526&7          &26.28  &1.090  &1.056  &13.1&11.7 ($<30.3$)      &$41.1$&$36.7$&$2.3~~(<95.1)$       &1.2 \\
4.199          &526&0          &26.55  &1.089  &1.056  &2.6&10.8 ($<21.0$)       &$8.1$&$33.6$&$0.4~~(<65.3)$        &$<1$ \\
4.209          &517&1          &26.40  &1.098  &1.057  &$-7.2$&9.6~ ($<14.2$)     &$-22.7$&$30.3$&$1.2~~(<44.8)$      &$<1$ \\
4.219          &514&6          &26.69  &1.099  &1.056  &11.6&10.2 ($<26.7$)      &$36.4$&$32.0$&$2.0~~(<83.7)$       &1.2 \\
4.226          &1100&9         &27.41  &1.101  &1.056  &136.1&20.7              &$193.9$&$29.5$&$10.7$             &7.8 \\
4.236          &530&3          &27.62  &1.105  &1.056  &16.7&10.6 ($<31.8$)      &$49.0$&$31.1$&$2.7~~(<93.4)$       &1.7 \\
4.244          &538&1          &27.72  &1.098  &1.056  &19.3&11.4 ($<35.3$)      &$55.8$&$33.0$&$3.1~~(102.1)$       &1.8 \\
4.258          &825&67         &27.98  &1.101  &1.054  &102.1&23.7              &$190.5$&$44.2$&$10.5$             &4.5 \\
4.267          &531&1          &28.08  &1.105  &1.053  &8.2&10.4 ($<24.3$)       &$23.6$&$30.0$&$1.3~~(<70.0)$       &$<1$ \\
4.278          &175&70         &27.86  &1.109  &1.053  &11.0&8.9~ ($<24.1$)       &$96.2$&$77.8$&$5.3~~(<210.7)$      &1.3 \\
4.288          &502&4          &28.05  &1.111  &1.053  &8.9&16.4 ($<34.4$)       &$27.0$&$49.8$&$1.5~~(<104.4)$      &$<1$ \\
4.312          &501&2          &28.59  &1.112  &1.052  &10.0&16.0 ($<34.8$)      &$29.8$&$47.7$&$1.6~~(<103.7)$      &$<1$ \\
4.337          &505&0          &29.09  &1.127  &1.051  &$-3.3$&18.1 ($<29.7$)    &$-9.5$&$52.0$&$0.5~~(<85.4)$       &$<1$ \\
4.377          &522&7          &29.67  &1.126  &1.051  &20.5&14.9 ($<41.5$)      &$55.8$&$40.6$&$3.1~~(<113.0)$      &1.4 \\
4.396          &507&8          &29.90  &0.969  &1.051  &13.1&14.6 ($<34.7$)      &$42.4$&$47.2$&$2.3~~(<112.3)$      &$<1$ \\
4.416          &1074&56        &30.19  &0.983  &1.052  &$-3.0$&18.8 ($<31.1$)    &$-4.5$&$28.0$&$0.2~~(<46.3)$       &$<1$ \\
4.436          &569&9          &30.34  &0.887  &1.054  &$-2.2$&15.5 ($<26.0$)    &$-6.8$&$48.0$&$0.4~~(<80.5)$       &$<1$ \\
4.467          &111&1          &30.27  &1.023  &1.055  &4.3&7.7 ~($<17.0$)        &$59.3$&$106.1$&$3.3~~(<234.3)$      &$<1$ \\
4.527          &112&12         &30.60  &1.029  &1.054  &17.6&7.6 ~($<28.2$)       &$236.5$&$102.1$&$13.0~(<378.9)$    &2.7 \\
4.599          &586&9          &31.05  &1.034  &1.055  &28.7&15.3 ($<49.7$)      &$72.2$&$38.5$&$4.0~~(<125.0)$      &2.0 \\
4.630          &521&53         &31.03  &1.045  &1.054  &7.7&14.7 ($<31.1$)       &$21.6$&$41.2$&$1.2~~(<87.2)$       &$<1$ \\
4.643          &551&65         &31.14  &1.054  &1.054  &16.1&17.7 ($<42.1$)      &$42.2$&$46.4$&$2.3~~(<110.3)$      &$<1$ \\
4.664          &529&43         &31.21  &1.060  &1.054  &$-8.8$&15.4 ($<22.6$)    &$-23.8$&$41.7$&$1.3~~(<61.2)$      &$<1$ \\
4.684          &1667&39        &31.32  &1.070  &1.054  &$-4.5$&28.7 ($<46.2$)    &$-3.8$&$24.4$&$0.2~~(<39.2)$       &$<1$ \\
4.701          &535&54         &31.17  &1.071  &1.055  &23.9&16.0 ($<46.2$)      &$63.4$&$42.4$&$3.5~~(<122.0)$      &1.6 \\
4.740          &163&87         &31.71  &1.015  &1.055  &5.8&10.1 ($<21.8$)       &$52.1$&$90.8$&$2.9~~(<195.9)$      &$<1$ \\
4.750          &366&55         &31.73  &1.025  &1.055  &$-5.7$&15.0 ($<25.8$)    &$-22.7$&$59.6$&$1.2~~(<102.6)$     &$<1$ \\
4.780          &511&47         &31.66  &1.036  &1.055  &45.7&17.2 ($<68.8$)      &$129.0$&$48.6$&$7.1~~(<194.2)$     &2.8 \\
4.843          &525&16         &31.73  &1.066  &1.056  &11.3&19.3 ($<40.9$)      &$30.1$&$51.5$&$1.7~~(<109.1)$      &$<1$ \\
4.918          &207&82         &31.53  &1.092  &1.056  &20.5&12.3 ($<37.7$)      &$135.7$&$81.4$&$7.5~~(<249.5)$     &1.8 \\
4.950          &159&28         &31.26  &1.100  &1.056  &$-4.3$&10.4 ($<16.6$)    &$-37.2$&$89.9$&$2.0~~(<143.5)$     &$<1$ \\
    \hline
    \end{tabular}}
    \label{num_result}
\end{table}
%
\section{Systematic uncertainty}
\noindent
The systematic uncertainty in the Born cross section measurements is evaluated for the following sources: the tracking and PID for charged particles, the vertex fit of $p K^- K^-$, the signal and background shapes, the input line shape and the quoted luminosity. Assuming all sources are independent, the total systematic uncertainty is calculated by the quadratic sum of the uncertainties due to all individual sources, as shown in table~\ref{SU}. All will be discussed in the following.
\subsection{Tracking}
\noindent
The systematic uncertainty associated with charged track reconstruction is evaluated as 1.0\% per track using the control samples of $J/\psi \to \pi^0 p \bar{p}$~\cite{tracking1} and $J/\psi \to K^* K$~\cite{tracking2}. Given that three charged tracks are reconstructed, the systematic uncertainty is set to be 3.0\%.
\subsection{PID}
\noindent
The systematic uncertainty due to the PID is evaluated as 1.0\% per track by the control sample of $J/\psi \to p \bar{p} \eta'$ and $\psi(3686) \to K^-\Lambda\bar{ \Xi}^+~+~$ c.c.~\cite{pid1, pid2}. There are three particles identified, so this systematic uncertainty is assigned as 3.0\%.
\subsection{Vertex fit}
\noindent
The systematic uncertainty for the vertex fit is evaluated by the control sample of $\psi(3686) \to \pi^+ \pi^- J/\psi$ with $J/\psi \to p \bar{p}$, where a proton or anti-proton is missed in the reconstruction. A cut for vertex $\chi^2<30$ is required to estimate the efficiency loss, and the difference for relative efficiency between data and MC sample is taken as the systematic uncertainty, which is assigned as 1.3\%.
\subsection{Signal and background shapes}
\noindent
Due to the limited statistics, this systematic uncertainty is estimated by combining all data samples.  The signal shape is changed from the MC shape convolved with a Gaussian function to the signal MC shape, and the background shape is adjusted from a second to third-order Chebyshev polynomial function. The differences in the fitted signal yields, 0.7\% for the variation of the signal and 3.0\% for the variation of the background shape, are taken as the systematic uncertainty.
\subsection{Input lineshape}
\noindent
The input lineshape is evaluated by varying the parameters that we used in the fit to the dressed cross section with the line shape of power-law function only(described in section~\ref{sec66}) in the iterative procedure. The fitted lineshape is needed to determine the ISR correction, but is only known up to the precision of the data. Thus, we use 200 lineshape samples drawn randomly from a multivariate Gaussian using the covariance matrix of the fit performed in section 5 to re-evaluate the ISR correction. The standard deviation of the product $\varepsilon \cdot (1+\delta_{\rm ISR})$ obtained in this way is used as a systematic uncertainty. This systematic uncertainty is negligible ($\sim 10^{-3}$).
\subsection{Luminosity}
\noindent
The integrated luminosity is estimated using Bhabha scattering events, and the systematic uncertainty for each energy point is assigned as 1\% \cite{lumi1,lumi2}.
\begin{table}[!htp]
    \centering
    \caption{Relative systematic uncertainties in the cross section measurements, where * marks represent the correlated systematic uncertainty.}
    \begin{tabular}{lc}
    \hline
    Source                              &Uncertainty~(\%)  \\
    \hline
    Tracking${}^*$                      &3.0 \\
    PID${}^*$                           &3.0 \\
    Vertex fit${}^*$                    &1.3 \\
    Signal and background shapes        &3.1 \\
    Input line shape                    &Negligible \\
    Luminosity${}^*$                    &1.0 \\
    \hline
    Total                               &5.5 \\
    \hline
    \end{tabular}
    \label{SU}
\end{table}

\section{Fit to the dressed cross section}
\label{sec66}
\begin{figure}[!htp]
    \label{UpperLimit}
    \centering
    \includegraphics[scale=0.35]{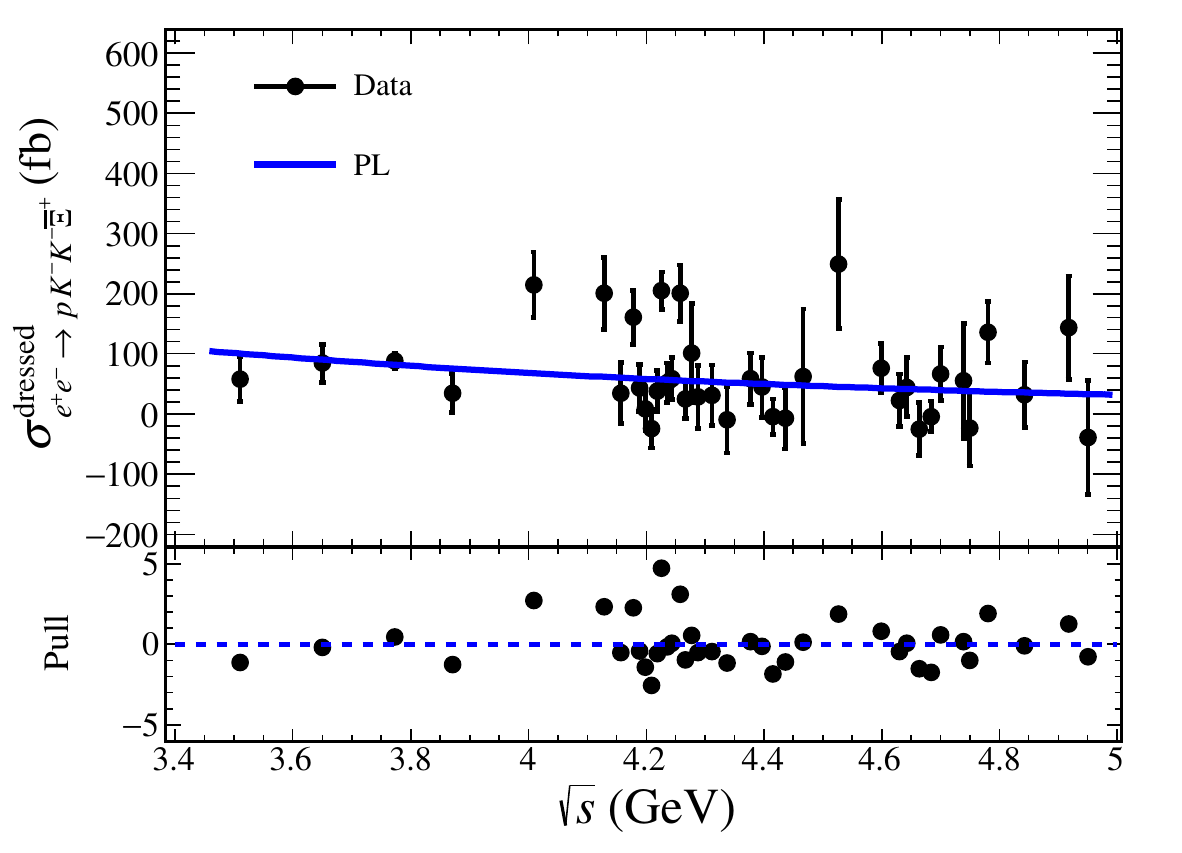}
    \includegraphics[scale=0.35]{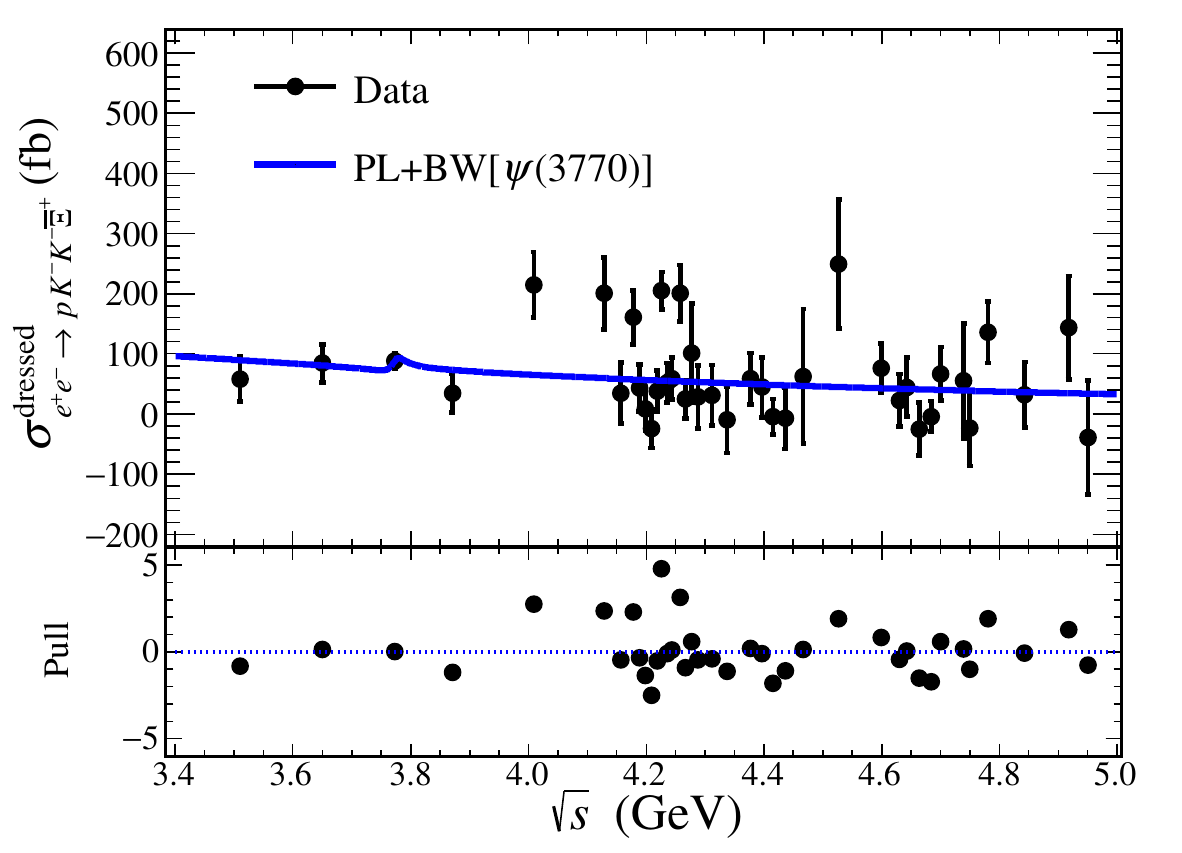}
    \includegraphics[scale=0.35]{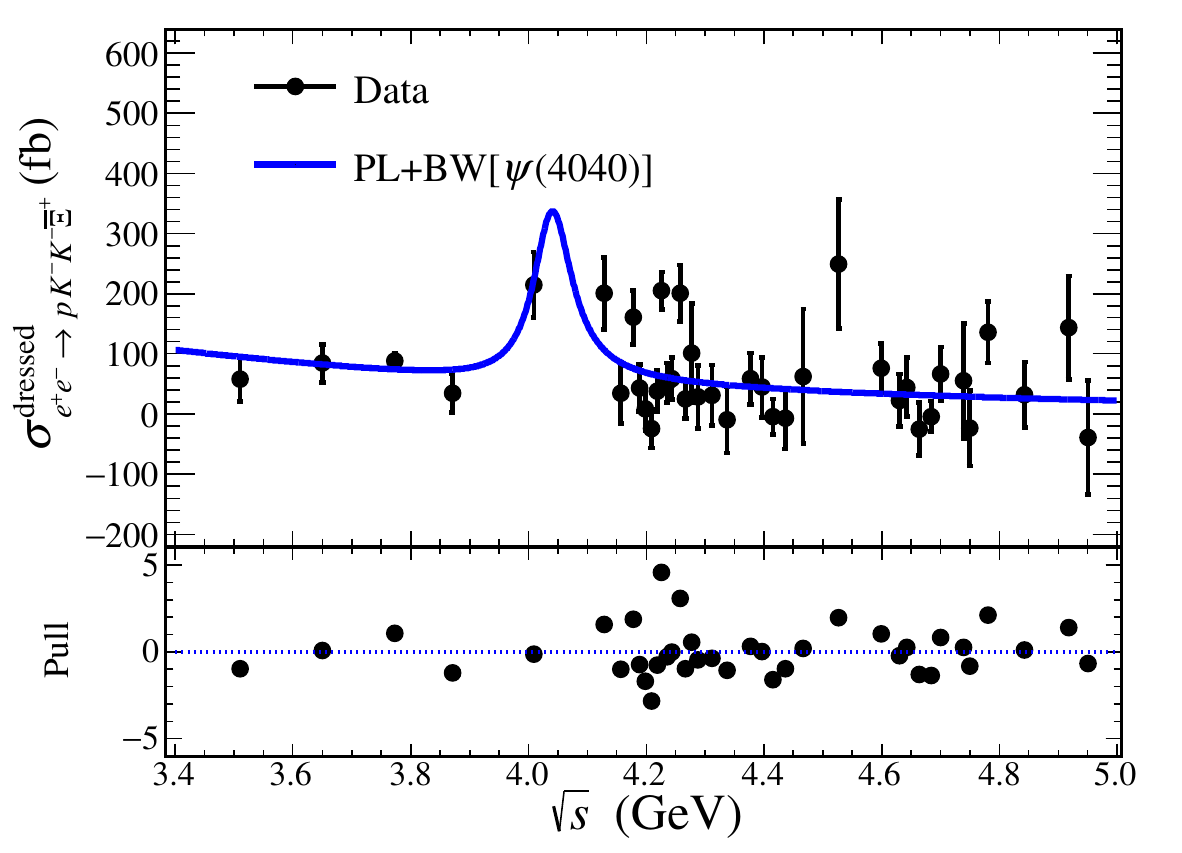}
    \includegraphics[scale=0.35]{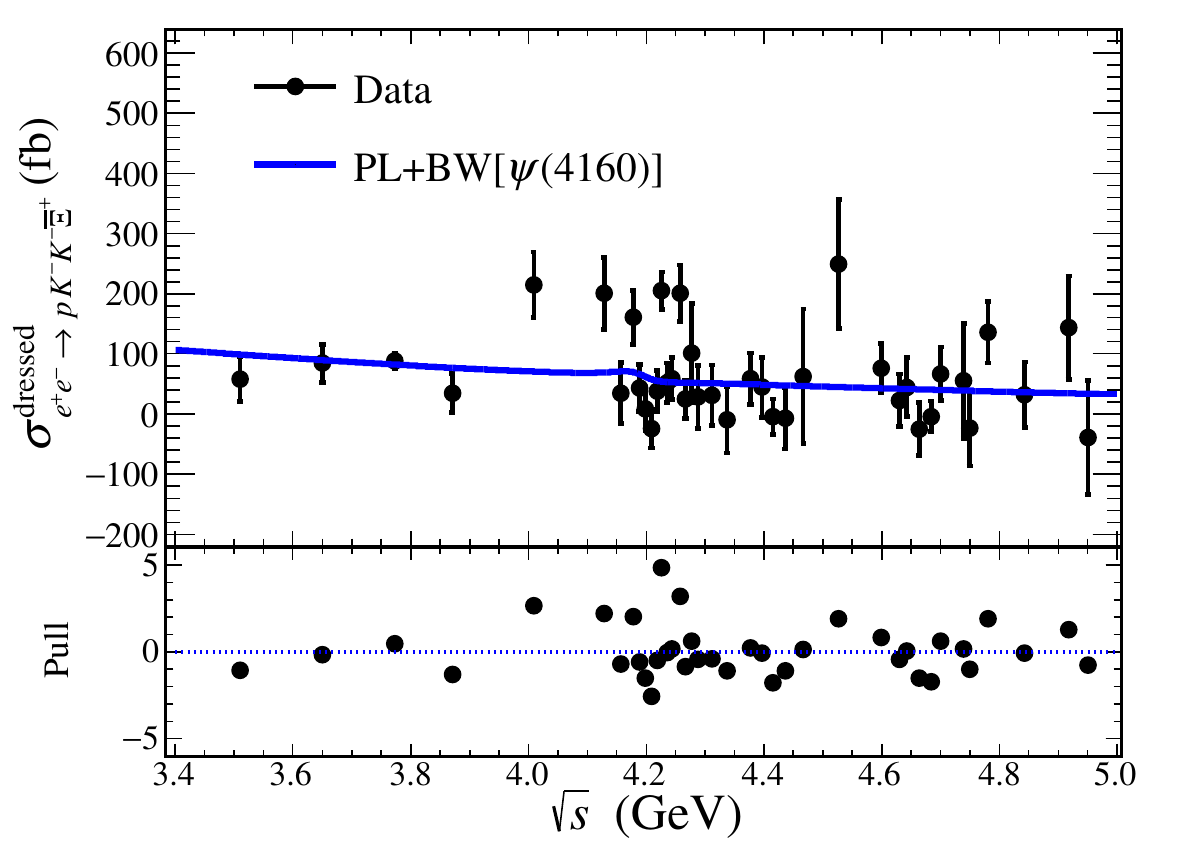}
    \includegraphics[scale=0.35]{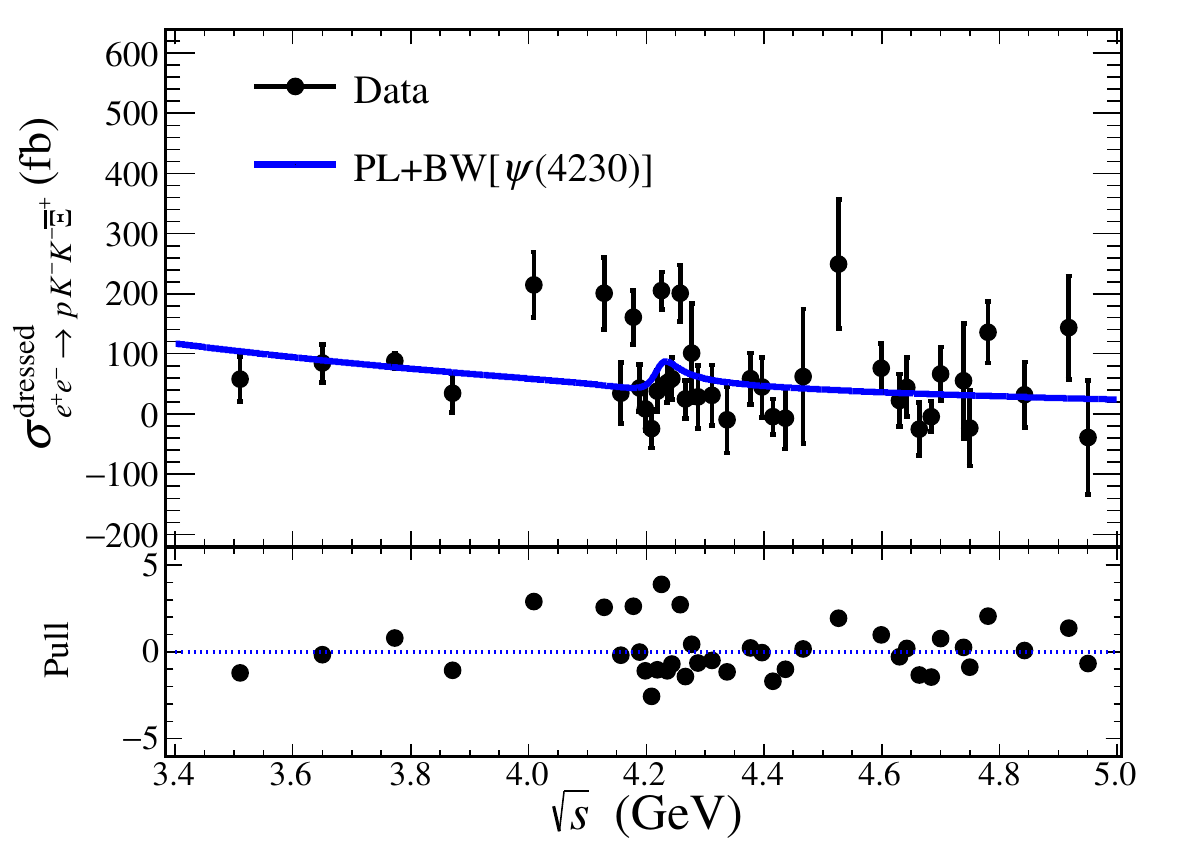}
    \includegraphics[scale=0.35]{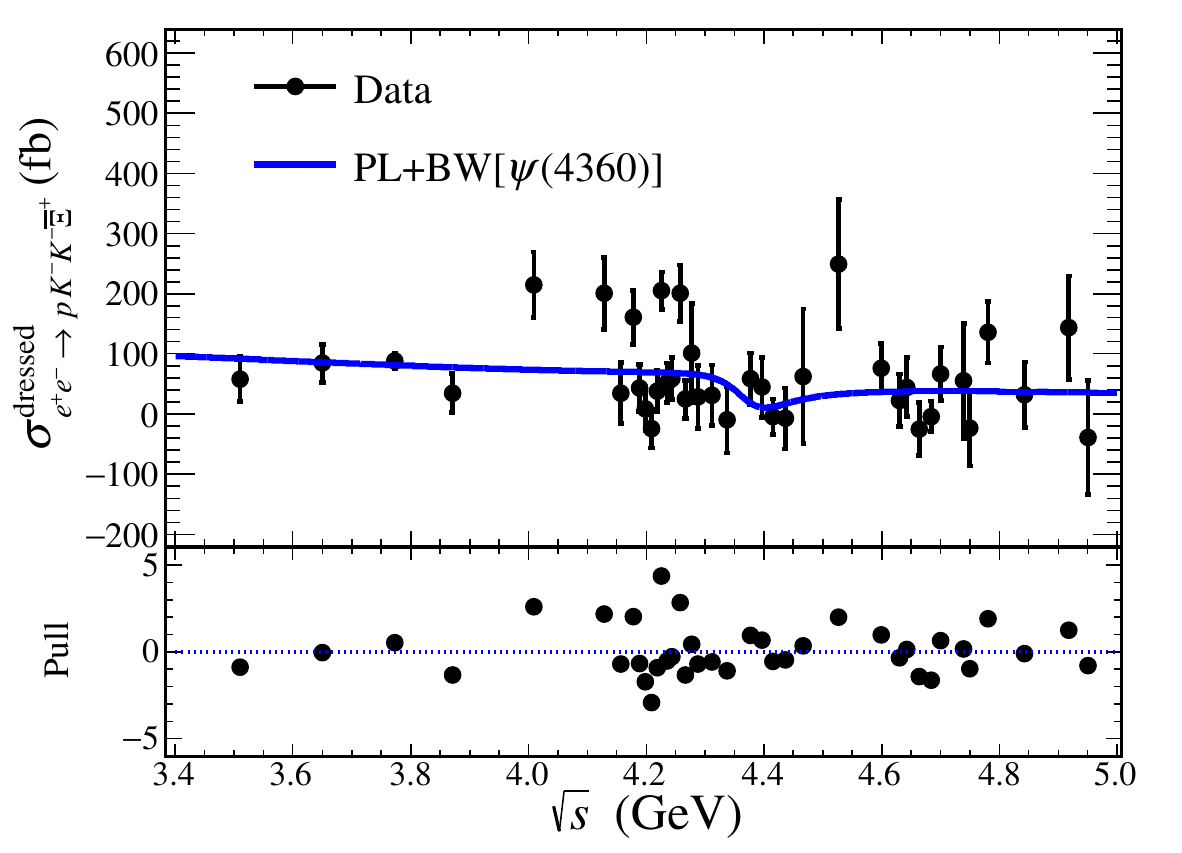}
    \includegraphics[scale=0.35]{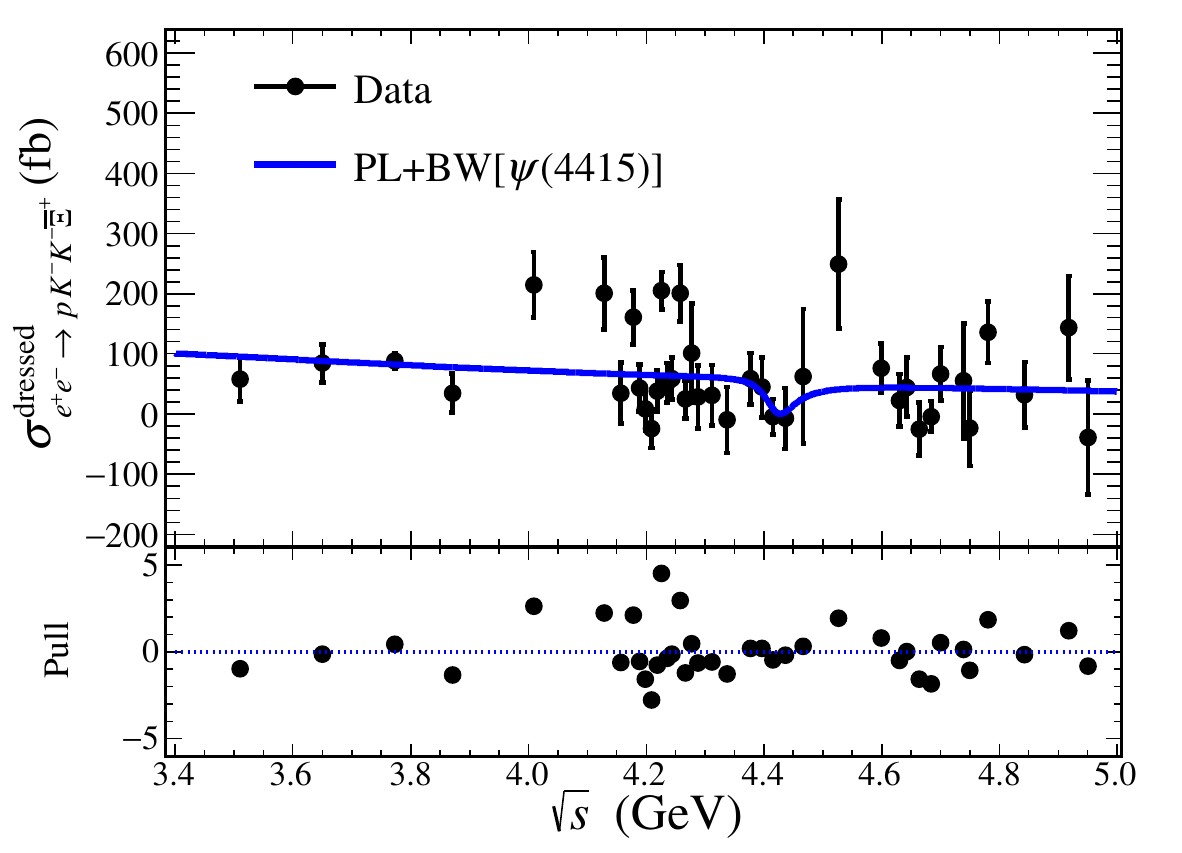}
    \includegraphics[scale=0.35]{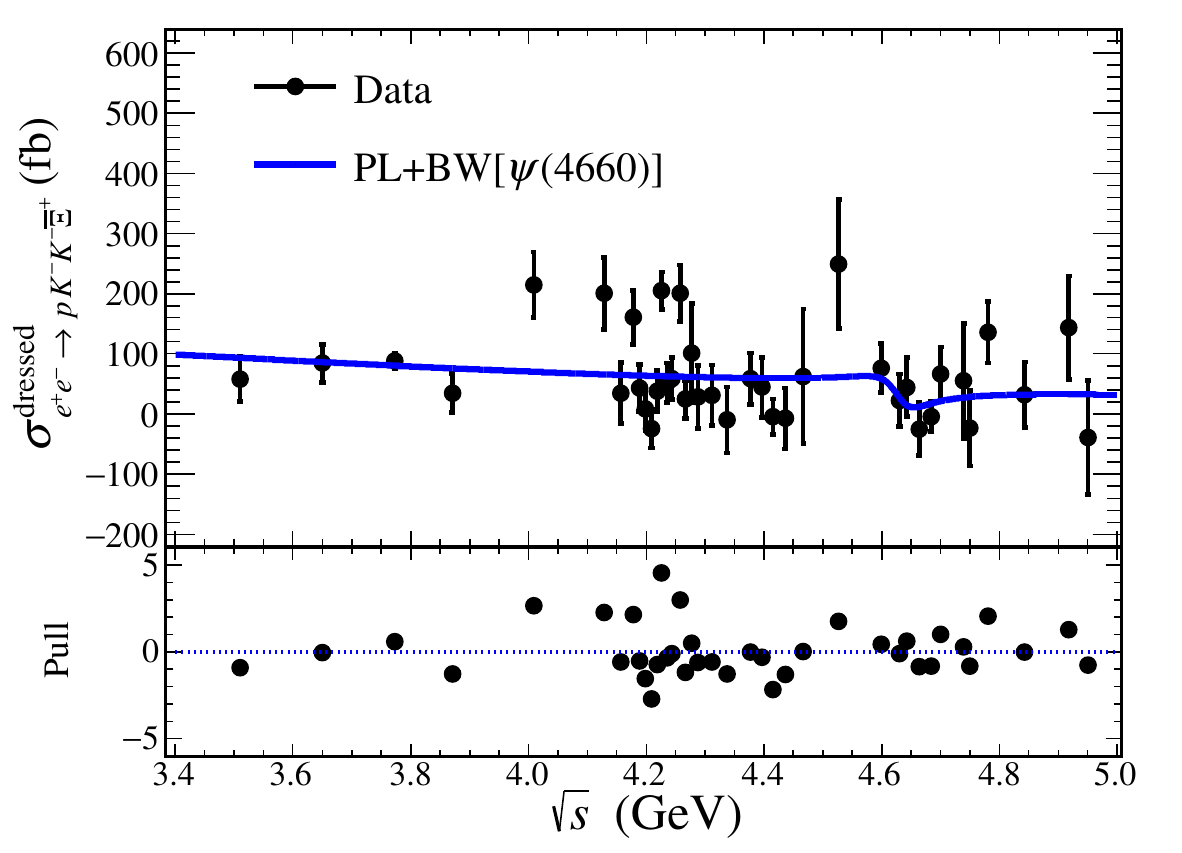}
    \caption{ Fits to the dressed cross sections under the assumption of the PL function only (top left), and the PL plus a resonance, i.e. $\psi(3770)$, $\psi(4040)$, $\psi(4160)$, $\psi(4230)$, $\psi(4360)$, $\psi(4415)$, or $\psi(4660)$. The black dots with error bars are the dressed cross sections, where the uncertainty is combined from both statistical and systematic uncertainties, and the blue lines denote the fit curves.  For each hypothesis, the panel below the main plot shows the pull distribution.}
     \label{fit_cross_section}
\end{figure}

\noindent
Potential resonances in the lineshape of the cross sections for the process $e^+ e^- \to p K^- K^- \bar{\Xi}$ are studied by fitting the dressed cross section $\sigma^{\rm dressed} = \frac{\sigma^{\rm B}}{\left| 1 - \Pi \right|^2}$ using the least $\chi^2$ method, where $\chi^2 = \Delta X^T V^{-1} \Delta X$, and $\Delta X$ is the vector of residuals between the measured and fitted cross section.
The covariance matrix $V$ incorporates the correlated and uncorrelated uncertainties among different energy points, where the systematic uncertainties for tracking, PID, vertex fit and luminosity mentioned in the previous section are assumed to be fully correlated if referring to the same energy point, and uncorrelated otherwise.
The line shape is described by a power-law (PL) function plus a Breit-Wigner (BW) function as
\begin{equation}
\label{bcs_function}
  \sigma(\sqrt{s}) \propto \left| c_0 P(\sqrt{s}) \frac{1}{\sqrt{s}^N} + e^{i \phi} {\rm BW}(\sqrt{s}; m, \Gamma, \Gamma_{ee} \mathcal{B})\right|^2.
\end{equation}
Here, $c_0$ and $N$ are free parameters.  When no intermediate resonances are included in the fit, we find $c_0 = (5.9 \pm 2.6) \times 10^{-5}$ and $N = 1.2 \pm 0.3$.
The $P(\sqrt{s})$ is the PHSP factor, $\phi$ is the relative phase, and the BW function is defined as
\begin{equation}
\label{bw}
  {\rm BW}(\sqrt{s};m,\Gamma,\Gamma_{ee} \mathcal{B}) = \frac{\sqrt{12 \pi \Gamma_{ee} \mathcal{B} \Gamma}}{s - m^2 + i m \Gamma},
\end{equation}
where $\Gamma_{ee} \mathcal{B}$ is the product of $e^+e^-$ partial decay width and branching fraction of the assumed resonance decaying into the $pK^-K^-\bar{\Xi}^+$ final state. In the current fit, both $\phi$ and $\Gamma_{ee} \mathcal{B}$ are the undetermined parameters.
The considered resonances are the $\psi(3770)$, $\psi(4040)$, $\psi(4160)$, $\psi(4230)$, $\psi(4360)$, $\psi(4415)$, and $\psi(4660)$, and their masses and widths are fixed to the values taken from the PDG~\cite{PDG2022}.
The significance of a resonance is determined from the change of $\chi^2$ and the number of degrees of freedom $n_{\rm dof}$ for the hypothesis with and without the resonance itself.
None of the considered resonances have a significance above $3\sigma$.
Here, the resonance parameters of $\psi(4160)$ and $\psi(4230)$ under the current statistics are hard to be exactly separated from each other because these two states are quite close and overlap and the interference picture can be sensitive to the resonance parameters.
Figure~\ref{fit_cross_section} shows the fit to the dressed cross sections without and with the different resonance assumptions.
The upper limits of $\Gamma_{ee} \mathcal{B}$ have been evaluated at the $90\%$ C.L. according to the $\chi^2$ distribution so that $\int_{0}^{\Delta \chi^2_{\rm UL}} \chi^2(x; \Delta n_{\rm df}) {\rm d}x = 90\%$, where $\Delta \chi^2_{\rm UL}$ is the difference of likelihood value between the evaluated nominal result and the upper limit one, $\Delta n_{\rm df}$ is the change in the degrees of freedom. The fit with only PL function presents a quality of $\chi^2/n_{\rm df} = 87/37$, corresponding to a $\chi^2$-probability around $10^{-6}$. The total numerical results are summarized in table~\ref{UpperFitPara}.
\begin{table}[!htp]
    \centering
    \caption{Results of the fits to the dressed cross sections. Here, the superscripts I and II denote the different solutions, the numbers in parentheses represent the nominal results, $S~(\sigma)$ is the significance for each resonance, and the upper limits of partial widths are evaluated at 90\% C.L.}
    \begin{tabular}{lr@{\hspace{0.2em}$\pm$\hspace{0.2em}}llcc}
    \hline
    Fit model                   &\multicolumn{2}{c}{~$\phi$ (rad)}      &$~\Gamma_{ee} \mathcal{B}$ ($\times 10^{-3}$ eV)     &$S~(\sigma)$      &$\chi^2/n_{\rm df}$ \\
    \hline
    PL                          &\multicolumn{2}{c}{~$-$}               &$~~~~~~~~~~~-$ &$-$        &$87/(39-2)$ \\
    \hline
    \quad$+\psi(3770)^{\rm I}$        &$0.57$&$4.70$                          &~~~~~~~~~~$(0.0\pm0.1)$             &\multirow{2}*{0.2}         &\multirow{2}*{$87/(39-4)$} \\
    \quad$+\psi(3770)^{\rm II}$       &$-1.52$&$0.39$                           &$<9.6~~~(8.6\pm0.7)$ \\
    \hline
    \quad$+\psi(4040)$                &$1.38$&$0.16$                          &$<17.8~(10.8\pm3.3)$     &2.8     &$76/(39-4)$ \\
    \hline
    \quad$+\psi(4160)^{\rm I}$        &$2.74$&$0.94$                          &~~~~~~~~~~$(0.0\pm0.1)$      &\multirow{2}*{0.1}   &\multirow{2}*{$87/(39-4)$} \\
    \quad$+\psi(4160)^{\rm II}$       &$-1.64$&$0.08$                         &$<25.4~(20.7\pm2.2)$ \\
    \hline
    \quad$+\psi(4230)^{\rm I}$        &$0.51$&$0.39$                          &~~~~~~~~~~$(0.5\pm0.3)$      &\multirow{2}*{1.5}     &\multirow{2}*{$83/(39-4)$} \\
    \quad$+\psi(4230)^{\rm II}$       &$1.42$&$0.08$                          &$<17.8~(14.9\pm1.3)$ \\
    \hline
    \quad$+\psi(4360)^{\rm I}$        &$-2.19$&$0.34$                         &~~~~~~~~~~$(3.4\pm3.4)$      &\multirow{2}*{1.7}     &\multirow{2}*{$83/(39-4)$} \\
    \quad$+\psi(4360)^{\rm II}$       &$-1.84$&$0.14$                         &$<33.3~(19.4\pm7.7)$ \\
    \hline
    \quad$+\psi(4415)$                &$-1.80$&$0.23$                         &$<14.7~(4.7\pm4.8)$      &1.7      &$82/(39-4)$ \\
    \hline
    \quad$+\psi(4660)^{\rm I}$        &$-2.46$&$0.48$                         &~~~~~~~~~~$(2.1\pm2.3)$              &\multirow{2}*{1.1}      &\multirow{2}*{$84/(39-4)$} \\
    \quad$+\psi(4660)^{\rm II}$       &$-1.90$&$0.20$                         &$<23.1~(13.1\pm5.3)$ \\
    \hline
    \end{tabular}
    \label{UpperFitPara}
\end{table}

\section{Summary}
\noindent
Using a data sample of $e^+ e^-$ collisions corresponding to a total integrated luminosity of 20 ${\rm fb}^{-1}$ collected with the BESIII detector at the BEPCII collider,
we present a measurement of the Born cross section for the process $e^+e^- \to p K^-K^-\bar{\Xi}^{+}$ at 39 center-of-mass energies between 3.5 and 4.9 GeV with a partial reconstruction technique.
A fit to the dressed cross sections for $e^+e^- \to p K^-K^-\bar{\Xi}^{+}$ under the assumption of one resonance, i.e. $\psi(3770)$, $\psi(4040)$, $\psi(4160)$, $\psi(4230)$, $\psi(4360)$, $\psi(4415)$ or $\psi(4660)$ plus a continuum contribution is performed.
The values of the fitted parameters for $\Gamma_{ee}\mathcal{B}$ for each assumed resonance are summarized in Table~\ref{UpperFitPara}. No statistically significant evidence for any of the considered resonances decaying into the $p K^-K^-\bar{\Xi}^{+}$ final state is found.
Upper limits on the product of $\Gamma_{ee}\mathcal{B}$ for all assumed resonances decaying into $p K^-K^-\bar{\Xi}^{+}$ are determined  at the $90\%$ C.L.
Compared with other processes~\cite{BESIII:2021ccp, BESIII:2023rse, BESIII:2024ogz}, there is still a lack of experimental information regarding the baryonic decays of charmonium(-like) states. The results of this work provide new experimental support on baryon production above the open charm region.

\acknowledgments
\noindent
The BESIII Collaboration thanks the staff of BEPCII (https://cstr.cn/31109.02.BEPC) and the IHEP computing center for their strong support. This work is supported in part by
the Fundamental Research Funds for the Central Universities Nos.
lzujbky-2025-ytA05, lzujbky-2025-it06, lzujbky-2024-jdzx06;
National Key R\&D Program of China under Contracts Nos. 2023YFA1606000, 2023YFA1606704; National Natural Science Foundation of China (NSFC) under Contracts Nos. 12247101, 11635010, 11735014, 11935015, 11935016, 11935018, 12025502, 12035009, 12035013, 12061131003, 12192260, 12192261, 12192262, 12192263, 12192264, 12192265, 12221005, 12225509, 12235017, 12361141819;
the Natural Science Foundation of Gansu Province Nos. 22JR5RA389, 25JRRA799;
the ‘111 Center’ under Grant No. B20063;
the Chinese Academy of Sciences (CAS) Large-Scale Scientific Facility Program; the CAS Center for Excellence in Particle Physics (CCEPP); Joint Large-Scale Scientific Facility Funds of the NSFC and CAS under Contract No. U1832207; CAS under Contract No. YSBR-101; 100 Talents Program of CAS; The Institute of Nuclear and Particle Physics (INPAC) and Shanghai Key Laboratory for Particle Physics and Cosmology; German Research Foundation DFG under Contract No. FOR5327; Istituto Nazionale di Fisica Nucleare, Italy; Knut and Alice Wallenberg Foundation under Contracts Nos. 2021.0174, 2021.0299; Ministry of Development of Turkey under Contract No. DPT2006K-120470; National Research Foundation of Korea under Contract No. NRF-2022R1A2C1092335; National Science and Technology fund of Mongolia; National Science Research and Innovation Fund (NSRF) via the Program Management Unit for Human Resources \& Institutional Development, Research and Innovation of Thailand under Contract No. B50G670107; Polish National Science Centre under Contract No. 2019/35/O/ST2/02907; Swedish Research Council under Contract No. 2019.04595; The Swedish Foundation for International Cooperation in Research and Higher Education under Contract No. CH2018-7756; U. S. Department of Energy under Contract No. DE-FG02-05ER41374.

\newpage
{\bf \noindent The BESIII collaboration}\\
\\
{\small
M.~Ablikim$^{1}$\BESIIIorcid{0000-0002-3935-619X},
M.~N.~Achasov$^{4,b}$\BESIIIorcid{0000-0002-9400-8622},
P.~Adlarson$^{77}$\BESIIIorcid{0000-0001-6280-3851},
X.~C.~Ai$^{82}$\BESIIIorcid{0000-0003-3856-2415},
R.~Aliberti$^{36}$\BESIIIorcid{0000-0003-3500-4012},
A.~Amoroso$^{76A,76C}$\BESIIIorcid{0000-0002-3095-8610},
Q.~An$^{73,59,\dagger}$,
Y.~Bai$^{58}$\BESIIIorcid{0000-0001-6593-5665},
O.~Bakina$^{37}$\BESIIIorcid{0009-0005-0719-7461},
Y.~Ban$^{47,g}$\BESIIIorcid{0000-0002-1912-0374}
% ...
\\
\vspace{0.2cm}
(BESIII Collaboration)\\
\vspace{0.2cm} {\it
$^{1}$ Institute of High Energy Physics, Beijing 100049, People's Republic of China\\
% ...
$^{4}$ Budker Institute of Nuclear Physics SB RAS (BINP), Novosibirsk 630090, Russia\\
% ...
$^{36}$ Johannes Gutenberg University of Mainz, Johann-Joachim-Becher-Weg 45, D-55099 Mainz, Germany\\
$^{37}$ Joint Institute for Nuclear Research, 141980 Dubna, Moscow region, Russia\\
% ...
$^{47}$ Peking University, Beijing 100871, People's Republic of China\\
% ...
$^{58}$ Southeast University, Nanjing 211100, People's Republic of China\\
$^{59}$ State Key Laboratory of Particle Detection and Electronics, Beijing 100049, Hefei 230026, People's Republic of China\\
% ...
$^{73}$ University of Science and Technology of China, Hefei 230026, People's Republic of China\\
% ...
$^{76}$ University of Turin and INFN, (A)University of Turin, I-10125, Turin, Italy; (B)University of Eastern Piedmont, I-15121, Alessandria, Italy; (C)INFN, I-10125, Turin, Italy\\
$^{77}$ Uppsala University, Box 516, SE-75120 Uppsala, Sweden\\
% ...
$^{82}$ Zhengzhou University, Zhengzhou 450001, People's Republic of China\\
\vspace{0.2cm}

$^{\dagger}$ Deceased\\
% ...
$^{b}$ Also at the Novosibirsk State University, Novosibirsk, 630090, Russia\\
% ...
$^{g}$ Also at State Key Laboratory of Nuclear Physics and Technology, Peking University, Beijing 100871, People's Republic of China\\
% ...
}}


\begin{thebibliography}{99}

\bibitem{ccbar1} N. Brambilla {\it et al.}, \textit{Heavy Quarkonium: Progress, Puzzles, and Opportunities}, \href{https://doi.org/10.1140/epjc/s10052-010-1534-9}{\textit{Eur. Phys. J. C} {\bf 71} (2011) 1534} [\href{https://arxiv.org/abs/1010.5827}{{arXiv:1010.5827}}] [\href{https://inspirehep.net/literature/874793}{\textsc{inSPIRE}}].

\bibitem{ccbar2} R. A. Briceno {\it et al.}, \textit{Issues and Opportunities in Exotic Hadrons}, \href{https://doi.org/10.1088/1674-1137/40/4/042001}{\textit{Chin. Phys. C} {\bf 40} (2016) 042001} [\href{https://arxiv.org/abs/1511.06779}{{arXiv:1511.06779}}] [\href{https://inspirehep.net/literature/1405969}{\textsc{inSPIRE}}].

\bibitem{qm} T. Barnes, S. Godfrey and E. S. Swanson, \textit{Higher charmonia}, \href{https://doi.org/10.1103/PhysRevD.72.054026}{\textit{Phys. Rev. D} {\bf 72} (2005) 054026} [\href{https://doi.org/10.48550/arXiv.hep-ph/0505002}{hep-ph/0505002}] [\href{https://inspirehep.net/literature/681760}{\textsc{inSPIRE}}].

\bibitem{besi} BES collaboration, \textit{Measurements of the cross section for $e^+ e^- \to$ hadrons at
 center-of-mass energies from 2 GeV to 5 GeV}, \href{https://doi.org/10.1103/PhysRevLett.88.101802}{\textit{Phys. Rev. Lett.} {\bf 88} (2002) 101802} [\href{https://arxiv.org/abs/hep-ex/0102003}{hep-ex/0102003}] [\href{https://inspirehep.net/literature/552757}{\textsc{inSPIRE}}].

\bibitem{belle1} \textsc{BaBar} collaboration, \textit{Observation of a broad structure in the $\pi^+ \pi^- J/\psi$ mass spectrum around 4.26 ${\rm GeV}/c^2$}, \href{https://doi.org/10.1103/PhysRevLett.95.142001}{\textit{Phys. Rev. Lett.} {\bf 95} (2005) 142001} [\href{https://arxiv.org/abs/hep-ex/0506081}{hep-ex/0506081}] [\href{https://inspirehep.net/literature/686354}{\textsc{inSPIRE}}].

\bibitem{belle2} B\textsc{elle} collaboration, \textit{Measurement of $e^+ e^- \to \pi^+ \pi^- J/\psi$ cross-section via initial state radiation at Belle}, \href{https://doi.org/10.1103/PhysRevLett.99.182004}{\textit{Phys. Rev. Lett.} {\bf 99} (2007) 182004} [\href{https://arxiv.org/abs/0707.2541}{{arXiv:0707.2541}}] [\href{https://inspirehep.net/literature/756012}{\textsc{inSPIRE}}].

\bibitem{belle3} \textsc{BaBar} collaboration, \textit{Evidence of a broad structure at an invariant mass of 4.32 ${\rm GeV}/c^2$ in the reaction $e^+ e^- \to \pi^+ \pi^- \psi(2S)$ Measured at BaBar}, \href{https://doi.org/10.1103/PhysRevLett.98.212001}{\textit{Phys. Rev. Lett.} {\bf 98} (2007) 212001} [\href{https://arxiv.org/abs/hep-ex/0610057}{hep-ex/0610057}] [\href{https://inspirehep.net/literature/729388}{\textsc{inSPIRE}}].

\bibitem{belle4} B\textsc{elle} collaboration, \textit{Observation of two resonant structures in $e^+ e^- \to \pi^+ \pi^- \psi(2S)$ via initial state radiation at Belle}, \href{https://doi.org/10.1103/PhysRevLett.99.142002}{\textit{Phys. Rev. Lett.} {\bf 99} (2007) 142002} [\href{https://arxiv.org/abs/0707.3699}{{arXiv:0707.3699}}] [\href{https://inspirehep.net/literature/756643}{\textsc{inSPIRE}}].

\bibitem{belle5} B\textsc{elle} collaboration, \textit{Observation of a near-threshold enhancement in the $e^+ e^- \to \Lambda^+_c \bar{\Lambda}^-_c$ cross section using initial-state radiation}, \href{https://doi.org/10.1103/PhysRevLett.101.172001}{\textit{Phys. Rev. Lett}. {\bf 101} (2008) 172001}, [\href{https://arxiv.org/abs/0807.4458}{{arXiv:0807.4458}}] [\href{https://inspirehep.net/literature/791660}{\textsc{inSPIRE}}].

\bibitem{belle6} B\textsc{elle} collaboration, \textit{Measurement of $e^+ e^- \to \pi^+ \pi^-\psi(2S)$ via Initial State Radiation at Belle}, \href{https://doi.org/10.1103/PhysRevD.91.112007}{\textit{Phys. Rev. D} {\bf 91} (2015) 112007} [\href{https://arxiv.org/abs/1410.7641}{{arXiv:1410.7641}}] [\href{https://inspirehep.net/literature/1324785}{\textsc{inSPIRE}}].

\bibitem{belle7} B\textsc{elle} collaboration, \textit{Study of $e^+ e^- \to \pi^+ \pi^- J/\psi$ and Observation of a Charged Charmoniumlike State at Belle}, \href{https://doi.org/10.1103/PhysRevLett.110.252002}{\textit{Phys. Rev. Lett.} {\bf 110} (2013) 252002} [\href{https://arxiv.org/abs/1304.0121}{{arXiv:1304.0121}}] [\href{https://inspirehep.net/literature/1225975}{\textsc{inSPIRE}}].

\bibitem{babar1} \textsc{BaBar} collaboration, \textit{Study of the reaction $e^{+}e^{-} \to \psi(2S) \pi^{+} \pi^{-}$ via initial-state radiation at BaBar}, \href{https://doi.org/10.1103/PhysRevD.89.111103}{\textit{Phys. Rev. D} {\bf 89} (2014) 111103} [\href{https://arxiv.org/abs/1211.6271}{{arXiv:1211.6271}}] [\href{https://inspirehep.net/literature/1204444}{\textsc{inSPIRE}}].

\bibitem{babar2} \textsc{BaBar} collaboration, \textit{Study of the reaction $e^+ e^- \to J/\psi \pi^+ \pi^-$ via initial-state radiation at BaBar}, \href{https://doi.org/10.1103/PhysRevD.86.051102}{\textit{Phys. Rev. D} {\bf 86} (2012) 051102(R)} [\href{https://arxiv.org/abs/1204.2158}{arXiv:1204.2158}] [\href{https://inspirehep.net/literature/1107905}{\textsc{inSPIRE}}].

\bibitem{Wang:2025dur}
X.~F.~Wang, X.~Liu and Y.~N.~Gao,
\textit{Hadron Production in Open-charm Meson Pair at $e^+e^-$ Collider},
[\href{https://arxiv.org/abs/2502.15117}{arXiv:2502.15117}]
[\textcolor{blue}{\href{https://inspirehep.net/literature/2893289}{\textsc{inSPIRE}}}].


\bibitem{cleo} CLEO collaboration, \textit{Charmonium decays of $Y(4260)$, $\psi(4160)$ and $\psi(4040)$}, \href{https://doi.org/10.1103/PhysRevLett.96.162003}{\textit{Phys. Rev. Lett.},  {\bf 96} (2006) 162003} [\href{https://arxiv.org/abs/hep-ex/0602034}{hep-ex/0602034}] [\href{https://inspirehep.net/literature/710864}{\textsc{inSPIRE}}].

\bibitem{bes1} BESIII collaboration, \textit{Study of $e^+ e^- \to \omega \chi_{cJ}$ at center-of-mass energies from 4.21 to 4.42 GeV}, \href{https://doi.org/10.1103/PhysRevLett.114.092003}{\textit{Phys. Rev. Lett.} {\bf 114} (2015) 092003} [\href{https://arxiv.org/abs/1410.6538}{arXiv:1410.6538}] [\href{https://inspirehep.net/literature/1323621}{\textsc{inSPIRE}}].

\bibitem{bes2} BESIII collaboration, \textit{Observation of three charmonium-like states with $J^{PC} = 1^{--}$ in $e^+ e^- \to D^{*0} D^{*-} \pi^+$}, \href{https://doi.org/10.1103/PhysRevLett.130.121901}{\textit{Phys. Rev. Lett.} {\bf 130} (2023) 121901} [\href{https://arxiv.org/abs/2301.07321}{{arXiv:2301.07321}}] [\href{https://inspirehep.net/literature/2645388}{\textsc{inSPIRE}}].

\bibitem{Close:2005iz} F. E. Close and P.R. Page, \textit{Gluonic charmonium resonances at \textsc{BaBar} and B\textsc{elle}?}, \href{https://doi.org/10.1016/j.physletb.2005.09.016}{\textit{Phys. Lett. B} {\bf 628} (2005) 215} [\href{https://arxiv.org/abs/hep-ph/0507199}{{hep-ph/0507199}}] [\href{https://inspirehep.net/literature/687628}{\textsc{inSPIRE}}].

\bibitem{Xia:2015mga} L. G. Xia, \textit{Study of $\psi(3770)$ decaying to baryon anti-baryon pairs}, \href{https://doi.org/10.1016/j.physletb.2016.03.015}{\textit{Phys. Lett. B} {\bf 756} (2016) 77} [\href{https://arxiv.org/abs/1508.02039}{{arXiv:1508.02039}}] [\href{https://inspirehep.net/literature/1387320}{\textsc{inSPIRE}}].

\bibitem{Chen:2016qju} H. X. Chen, W. Chen, X. Liu and S. L. Zhu, \textit{The hidden-charm pentaquark and tetraquark states}, \href{https://doi.org/10.1016/j.physrep.2016.05.004}{\textit{Phys. Rept.} {\bf 639} (2016) 1} [\href{https://arxiv.org/abs/1601.02092}{{arXiv:1601.02092}}] [\href{https://inspirehep.net/literature/1414795}{\textsc{inSPIRE}}].

\bibitem{Qian:2021neg} R. Q. Qian, Q. Huang and X. Liu, \textit{Predicted $\Lambda\bar{\Lambda}$ and $\Xi^-\bar{\Xi}^+$ decay modes of the charmoniumlike $Y(4230)$}, \href{https://doi.org/10.1016/j.physletb.2022.137292}{\textit{Phys. Lett. B} {\bf 833} (2022) 137292} [\href{https://arxiv.org/abs/2111.13821}{{arXiv:2111.13821}}] [\href{https://inspirehep.net/literature/1978833}{\textsc{inSPIRE}}].

\bibitem{Bai:2023dhc} Z. Y. Bai, Q. S. Zhou and X. Liu, \textit{Higher strangeonium decays into light flavor baryon pairs like $\Lambda\bar{\Lambda}$, $\Sigma\bar{\Sigma}$, and $\Xi\bar{\Xi}$}, \href{https://doi.org/10.1103/PhysRevD.108.094036}{\textit{Phys. Rev. D } {\bf 108} (2023) 094036} [\href{https://arxiv.org/abs/2307.16255}{{arXiv:2307.16255}}] [\href{https://inspirehep.net/literature/2683598}{\textsc{inSPIRE}}].

\bibitem{Yan:2023yff} B. Yan, C. Chen and J. J. Xie, \textit{$\Sigma$ and $\Xi$ electromagnetic form factors in the extended vector meson dominance model}, \href{https://doi.org/10.1103/PhysRevD.107.076008}{\textit{Phys. Rev. D} {\bf 107} (2023) 076008} [\href{https://arxiv.org/abs/2301.00976}{{arXiv:2301.00976}}] [\href{https://inspirehep.net/literature/2620121}{\textsc{inSPIRE}}].

\bibitem{Dai:2023vsw} J. P. Dai, X. Cao and H. Lenske, \textit{Data driven isospin analysis of timelike octet baryons electromagnetic form factors and charmonium decay into baryon-anti-baryon}, \href{https://doi.org/10.1016/j.physletb.2023.138192}{\textit{Phys. Lett. B} {\bf 846} (2023) 138192} [\href{https://arxiv.org/abs/2304.04913}{{arXiv:2304.04913}}] [\href{https://inspirehep.net/literature/2650466}{\textsc{inSPIRE}}].

\bibitem{Yuan:2021wpg} C. Z. Yuan, \textit{Charmonium and charmoniumlike states at the BESIII experiment}, \href{https://doi.org/10.1093/nsr/nwab182}{\textit{Natl. Sci. Rev.} {\bf 8} (2021) nwab182} [\href{https://arxiv.org/abs/2102.12044}{{arXiv:2102.12044}}] [\href{https://inspirehep.net/literature/1848217}{\textsc{inSPIRE}}].






\bibitem{Ablikim:2013pgf}
BESIII collaboration,
\textit{Search for baryonic decays of $\psi(3770)$ and $\psi(4040)$},
\textcolor{blue}{\href{https://doi.org/10.1103/PhysRevD.87.112011}{\textit{Phys. Rev. D} \textbf{87} (2013) 112011}}
[\textcolor{blue}{\href{https://arxiv.org/abs/1305.1782}{arXiv:1305.1782}}]
[\textcolor{blue}{\href{https://inspirehep.net/literature/1232386}{\textsc{inSPIRE}}}].


\bibitem{BESIII:2016ssr}
BESIII collaboration,
\textit{Study of $\psi$ decays to the $\Xi^{-}\bar\Xi^{+}$ and $\Sigma(1385)^{\mp}\bar\Sigma(1385)^{\pm}$ final states},
\textcolor{blue}{\href{https://doi.org/10.1103/PhysRevD.93.072003}{\textit{Phys. Rev. D} \textbf{93} (2016) 072003}}
[\textcolor{blue}{\href{https://arxiv.org/abs/1602.06754}{arXiv:1602.06754}}]
[\textcolor{blue}{\href{https://inspirehep.net/literature/1422780}{\textsc{inSPIRE}}}].

\bibitem{BESIII:2016nix}
BESIII collaboration,
\textit{Study of $J/\psi$ and $\psi(3686)\rightarrow\Sigma(1385)^{0}\bar\Sigma(1385)^{0}$ and $\Xi^0\bar\Xi^{0}$},
\textcolor{blue}{\href{https://doi.org/10.1016/j.physletb.2017.03.046}{\textit{Phys. Lett. B} \textbf{770} (2017) 217}}
[\textcolor{blue}{\href{https://arxiv.org/abs/1607.06446}{arXiv:1607.06446}}]
[\textcolor{blue}{\href{https://inspirehep.net/literature/1477405}{\textsc{inSPIRE}}}].


\bibitem{Wang:2018kdh}
X.~F.~Wang, B.~Li, Y.~N.~Gao and X.~Lou,
\textit{Helicity amplitude analysis of $J/\psi$ and $\psi(3686)\to\Xi(1530)\bar{\Xi}(1530)$},
\textcolor{blue}{\href{https://www.sciencedirect.com/science/article/pii/S0550321319300689?via\%3Dihub}{\textit{Nucl. Phys. B} \textbf{941} (2019) 861}}
[\textcolor{blue}{\href{https://arxiv.org/abs/1811.11352}{arXiv:1811.11352}}]
[\textcolor{blue}{\href{https://inspirehep.net/literature/1694816}{\textsc{inSPIRE}}}].


\bibitem{BESIII:2019dve}
BESIII collaboration,
\textit{Observation of $\psi(3686)\rightarrow\Xi(1530)^{-}\bar\Xi(1530)^{+}$ and $\Xi(1530)^{-}\bar\Xi^{+}$},
\textcolor{blue}{\href{https://doi.org/10.1103/PhysRevD.100.051101}{\textit{Phys. Rev. D} \textbf{100} (2019) 051101}}
[\textcolor{blue}{\href{https://arxiv.org/abs/1907.13041}{arXiv:1907.13041}}]
[\textcolor{blue}{\href{https://inspirehep.net/literature/1747092}{\textsc{inSPIRE}}}].

\bibitem{Ablikim:2019kkp}
BESIII collaboration,
\textit{Measurement of the cross section for $e^{+}e^{-}\rightarrow\Xi^{-}\bar\Xi^{+}$ and observation of an excited $\Xi$ baryon},
\textcolor{blue}{\href{https://doi.org/10.1103/PhysRevLett.124.032002}{\textit{Phys. Rev. Lett.} \textbf{124} (2020) 032002}}
[\textcolor{blue}{\href{https://arxiv.org/abs/1910.04921}{arXiv:1910.04921}}]
[\textcolor{blue}{\href{https://inspirehep.net/literature/1758883}{\textsc{inSPIRE}}}].

\bibitem{BESIII:2020ktn}
BESIII collaboration,
\textit{Measurement of cross section for $e^+e^-\to\Xi^-\bar{\Xi}^+$ near threshold at BESIII},
\textcolor{blue}{\href{https://doi.org/10.1103/PhysRevD.103.012005}{\textit{Phys. Rev. D} \textbf{103} (2021) 012005}}
[\textcolor{blue}{\href{https://arxiv.org/abs/2010.08320}{arXiv:2010.08320}}]
[\textcolor{blue}{\href{https://inspirehep.net/literature/1823448}{\textsc{inSPIRE}}}].

\bibitem{BESIII:2021aer}
BESIII collaboration,
\textit{Measurement of cross section for $e^{+}e^{-}\rightarrow\Xi^{0}\bar{\Xi}^{0}$ near threshold},
\textcolor{blue}{\href{https://doi.org/10.1016/j.physletb.2021.136557}{\textit{Phys. Lett. B} \textbf{820} (2021) 136557}}
[\textcolor{blue}{\href{https://arxiv.org/abs/2105.14657}{arXiv:2105.14657}}]
[\textcolor{blue}{\href{https://inspirehep.net/literature/1866233}{\textsc{inSPIRE}}}].

\bibitem{BESIII:2021ccp}
BESIII collaboration,
\textit{Measurement of the cross section for $e^{+}e^{-}\rightarrow\Lambda\bar\Lambda$~and evidence of the decay $\psi(3770)\rightarrow\Lambda\bar\Lambda$},
\textcolor{blue}{\href{https://doi.org/10.1103/PhysRevD.104.L091104}{\textit{Phys. Rev. D} \textbf{104} (2021) L091104}}
[\textcolor{blue}{\href{https://arxiv.org/abs/2108.02410}{arXiv:2108.02410}}]
[\textcolor{blue}{\href{https://inspirehep.net/literature/1900124}{\textsc{inSPIRE}}}].

\bibitem{BESIII:2021gca}
BESIII collaboration,
\textit{Observation of $\psi(3686)\to\Xi(1530)^{0}\bar{\Xi}(1530)^{0}$ and $\Xi(1530)^{0}\bar{\Xi}^0$},
\textcolor{blue}{\href{https://doi.org/10.1103/PhysRevD.104.092012}{\textit{Phys. Rev. D} \textbf{104} (2021) 092012}}
[\textcolor{blue}{\href{https://arxiv.org/abs/2109.06621}{arXiv:2109.06621}}]
[\textcolor{blue}{\href{https://inspirehep.net/literature/1921775}{\textsc{inSPIRE}}}].

\bibitem{BESIII:2021cvv}
BESIII collaboration,
\textit{Measurement of $\Lambda$ baryon polarization in $e^+e^-\rightarrow\Lambda\bar\Lambda$ at $\sqrt{s} = 3.773$ GeV},
\href{https://doi.org/10.1103/PhysRevD.105.L011101}{\textit{Phys. Rev. D} \textbf{105} (2022) L011101 }
[\href{https://arxiv.org/abs/2111.11742}{arXiv:2111.11742}]
[\textcolor{blue}{\href{https://inspirehep.net/literature/1974025}{\textsc{inSPIRE}}}].

\bibitem{Wang:2022zyc}
X.~F.~Wang and G.~S.~Huang,
\textit{Electromagnetic Form Factor of Doubly-Strange Hyperon},
\href{https://doi.org/10.3390/sym14010065}{\textit{Symmetry} \textbf{14} (2022) 65}
[\textcolor{blue}{\href{https://inspirehep.net/literature/2037456}{\textsc{inSPIRE}}}].

\bibitem{BESIII:2022mfx}
BESIII collaboration,
\textit{Study of the processes $\chi_{cJ} \to \Xi^- \overline{\Xi}^+$ and $\Xi^0 \overline{\Xi}^0$},
\href{https://doi.org/10.1007/JHEP06(2022)074}{\textit{JHEP} \textbf{06} (2022) 074}
[\href{https://arxiv.org/abs/2202.08058}{arXiv:2202.08058}]
[\textcolor{blue}{\href{https://inspirehep.net/literature/2033855}{\textsc{inSPIRE}}}].

\bibitem{BESIII:2022lsz}
BESIII collaboration,
\textit{Observation of $\Xi^-$ hyperon transverse polarization in $\psi(3686) \to \Xi^-\overline{\Xi}^+$},
\href{https://doi.org/10.1103/PhysRevD.106.L091101}{\textit{Phys. Rev. D} \textbf{106} (2022) L091101}
[\href{https://arxiv.org/abs/2206.10900}{arXiv:2206.10900}]
[\textcolor{blue}{\href{https://inspirehep.net/literature/2099144}{\textsc{inSPIRE}}}].

\bibitem{BESIII:2022kzc}
BESIII collaboration,
\textit{Study of $e^+e^- \to \Omega^-\overline{\Omega}^+$ at center-of-mass energies from 3.49 to 3.67~GeV},
\href{https://doi.org/10.1103/PhysRevD.107.052003}{\textit{Phys. Rev. D} \textbf{107} (2023) 052003}
[\href{https://arxiv.org/abs/2212.03693}{arXiv:2212.03693}]
[\textcolor{blue}{\href{https://inspirehep.net/literature/2611486}{\textsc{inSPIRE}}}].

\bibitem{BESIII:2023lkg}
BESIII collaboration,
\textit{First simultaneous measurement of $\Xi^0$ and $\overline{\Xi}^0$ asymmetry parameters in $\psi(3686)$ decay},
\href{https://doi.org/10.1103/PhysRevD.108.L011101}{\textit{Phys. Rev. D} \textbf{108} (2023) L011101}
[\href{https://arxiv.org/abs/2302.09767}{arXiv:2302.09767}]
[\textcolor{blue}{\href{https://inspirehep.net/literature/2634735}{\textsc{inSPIRE}}}].

\bibitem{Liu:2023xhg}
H.~Liu, J.~Zhang and X.~Wang,
\textit{$CP$ Asymmetry in the $\Xi$ Hyperon Sector},
\href{https://doi.org/10.3390/sym15010214}{\textit{Symmetry} \textbf{15}(2023) 214}
[\textcolor{blue}{\href{https://inspirehep.net/literature/2633221}{\textsc{inSPIRE}}}].

\bibitem{BESIII:2023euh}
BESIII collaboration,
\textit{Measurement of $\Lambda$ transverse polarization in $e^+e^-$ collisions at $\sqrt{s} = 3.68$--$3.71$ GeV},
\href{https://doi.org/10.1007/JHEP10(2023)081}{\textit{JHEP} \textbf{10} (2023) 081}
[\href{https://arxiv.org/abs/2303.00271}{arXiv:2303.00271}]
[\textcolor{blue}{\href{https://inspirehep.net/literature/2637702}{\textsc{inSPIRE}}}].


\bibitem{BESIII:2023rwv}
BESIII collaboration,
\textit{Measurement of Energy-Dependent Pair-Production Cross Section and Electromagnetic Form Factors of a Charmed Baryon},
\href{https://doi.org/10.1103/PhysRevLett.131.191901}{\textit{Phys. Rev. Lett.} \textbf{131} (2023) 191901}
[\href{https://arxiv.org/abs/2307.07316}{arXiv:2307.07316}]
[\textcolor{blue}{\href{https://inspirehep.net/literature/2677290}{\textsc{inSPIRE}}}].

\bibitem{BESIII:2023rse}
BESIII collaboration,
\textit{Measurement of the cross section of $e^+e^-\to \Xi^-\overline{\Xi}^+$ at center-of-mass energies between 3.510 and 4.843 GeV},
\href{https://doi.org/10.1007/JHEP11(2023)228}{\textit{JHEP} \textbf{11} (2023) 228 }
[\href{https://arxiv.org/abs/2309.04215}{arXiv:2309.04215}]
[\textcolor{blue}{\href{https://inspirehep.net/literature/2695411}{\textsc{inSPIRE}}}].

\bibitem{BESIII:2024ogz}
BESIII collaboration,
\textit{Measurement of the cross sections of $e^+e^-\to K^-\overline{\Xi}^+\Lambda/\Sigma^0$ at center-of-mass energies between 3.510 and 4.914 GeV},
\href{https://doi.org/10.1007/JHEP07(2024)258}{\textit{JHEP} \textbf{07} (2024) 258}
[\href{https://arxiv.org/abs/2406.18183}{arXiv:2406.18183}]
[\textcolor{blue}{\href{https://inspirehep.net/literature/2802333}{\textsc{inSPIRE}}}].

\bibitem{BESIII:2024dmr}
BESIII collaboration,
\textit{Measurement of $\Sigma^+$ transverse polarization in $e^+e^-$ collisions at $\sqrt{s} = 3.68$--$3.71$ GeV},
\href{https://doi.org/10.1007/JHEP12(2024)186}{\textit{JHEP} \textbf{12} (2024) 186}
[\href{https://arxiv.org/abs/2408.03205}{arXiv:2408.03205}]
[\textcolor{blue}{\href{https://inspirehep.net/literature/2815316}{\textsc{inSPIRE}}}].

\bibitem{BESIII:2024ues}
BESIII collaboration,
\textit{Measurement of Born cross sections of $e^+e^-\to \Xi^0\overline{\Xi}^0$ and search for charmonium(-like) states at $\sqrt{s} = 3.51$--$4.95$ GeV},
\href{https://doi.org/10.1007/JHEP11(2024)062}{\textit{JHEP} \textbf{11} (2024) 062}
[\href{https://arxiv.org/abs/2409.00427}{arXiv:2409.00427}]
[\textcolor{blue}{\href{https://inspirehep.net/literature/2824143}{\textsc{inSPIRE}}}].

\bibitem{BESIII:2024umc}
BESIII collaboration,
\textit{Measurement of Born cross section of $e^+e^-\to \Sigma^+\overline{\Sigma}^-$ at center-of-mass energies between 3.510 and 4.951 GeV},
\href{https://doi.org/10.1007/JHEP05(2024)022}{\textit{JHEP} \textbf{05} (2024) 022}
[\href{https://arxiv.org/abs/2401.09468}{arXiv:2401.09468}]
[\textcolor{blue}{\href{https://inspirehep.net/literature/2748736}{\textsc{inSPIRE}}}].


\bibitem{BESIII:2024gql}
BESIII collaboration,
\textit{Measurement of Born cross section of $e^+e^-\to \Sigma^0\overline{\Sigma}^0$ at $\sqrt{s} = 3.50$--$4.95$ GeV},
\href{s://doi.org/10.1103/PhysRevD.108.L011101}{\textit{Phys. Rev. D} \textbf{111}  (2025) L051502}
[\href{https://arxiv.org/abs/2412.20305}{arXiv:2412.20305}]
[\textcolor{blue}{\href{https://inspirehep.net/literature/2863767}{\textsc{inSPIRE}}}].



\bibitem{BESIII:2025yzk}
BESIII collaboration,
\textit{Observation of Transverse Polarization and Determination of Electromagnetic Form Factor of $\Lambda$ Hyperon at $\sqrt{s}= 3.773$ GeV},
[\href{https://arxiv.org/abs/2504.05584}{arXiv:2504.05584}]
[\textcolor{blue}{\href{https://inspirehep.net/literature/2909351}{\textsc{inSPIRE}}}].


\bibitem{Zhang:2025qmo}
R.~Zhang and X.~Wang,
\textit{Search for Charmonium(-like) states decaying into the $\Omega^-\bar\Omega^+$ final states},
[\href{https://arxiv.org/abs/2508.03454}{arXiv:2508.03454}]
[\textcolor{blue}{\href{https://inspirehep.net/literature/2957692}{\textsc{inSPIRE}}}].







\bibitem{lumi1} BESIII collaboration, \textit{Precision measurement of the integrated luminosity of the data taken by BESIII at center of mass energies between 3.810 GeV and 4.600 GeV}, \href{https://doi.org/10.1088/1674-1137/39/9/093001}{\textit{Chin. Phys. C} {\bf 39} (2015) 093001} [\href{https://arxiv.org/abs/1503.03408}{{arXiv:1503.03408}}] [\href{https://inspirehep.net/literature/1351765}{\textsc{inSPIRE}}].

\bibitem{lumi2} BESIII collaboration, \textit{Luminosities and energies of $e^+ e^-$ collision data taken between = $4.61$ GeV and $4.95$ GeV at BESIII*}, \href{https://doi.org/10.1088/1674-1137/ac84cc}{\textit{Chin. Phys. C} \textbf{46} (2022) 113003} [\href{https://arxiv.org/abs/2205.04809}{{arXiv:2205.04809}}] [\href{https://inspirehep.net/literature/2079606}{\textsc{inSPIRE}}].

\bibitem{besiii1} BESIII collaboration, \textit{Design and Construction of the BESIII Detector}, \href{https://doi.org/10.1016/j.nima.2009.12.050}{\textit{Nucl. Instrum. Methods Phys. Res., Sect. A} {\bf 614} (2010) 345} [\href{https://arxiv.org/abs/0911.4960}{arXiv:0911.4960}] [\href{https://inspirehep.net/literature/838149}{\textsc{inSPIRE}}].

\bibitem{besiii2} BESIII collaboration, \textit{Future Physics Programme of BESIII}, \href{https://doi.org/10.1088/1674-1137/44/4/040001}{\textit{Chin. Phys. C} \textbf{44} (2020) 040001}[\href{https://arxiv.org/abs/1912.05983}{arXiv:1912.05983}][\href{https://inspirehep.net/literature/1770442}{\textsc{inSPIRE}}].

\bibitem{Yu:IPAC2016-TUYA01} C.~H.~Yu {\it et al.}, \textit{BEPCII Performance and Beam Dynamics Studies on Luminosity},
  Proceedings of IPAC2016, Busan, Korea, 2016, \href{https://doi.org/10.18429/JACoW-IPAC2016-TUYA01}{10.18429/JACoW-IPAC2016-TUYA01}.

\bibitem{etof1}
 X. Li {\it et al.}, \textit{Study of MRPC technology for BESIII endcap-TOF upgrade}, \href{https://doi.org/10.1007/s41605-017-0014-2}{\textit{Radiat. Detect. Technol. Methods} {\bf 1} (2017) 13} [\href{https://inspirehep.net/literature/2731162}{\textsc{inSPIRE}}].

 \bibitem{etof2}
 Y. X. Guo {\it et al.}, \textit{The study of time calibration for upgraded end cap TOF of BESIII}, \href{https://doi.org/10.1007/s41605-017-0012-4}{\textit{Radiat. Detect. Technol. Methods} {\bf 1} (2017) 15} [\href{https://inspirehep.net/literature/2731166}{\textsc{inSPIRE}}].

 \bibitem{etof3}
 P. Cao {\it et al.}, \textit{Design and construction of the new BESIII endcap Time-of-Flight system with MRPC Technology}, \href{https://doi.org/10.1016/j.nima.2019.163053}{\textit{Nucl. Instrum. Meth. A} {\bf 953} (2020) 163053} [\href{https://inspirehep.net/literature/1775466}{\textsc{inSPIRE}}].

\bibitem{geant41} \textsc{GEANT4 collaboration}, \textit{GEANT4-a simulation toolkit}, \href{https://doi.org/10.1016/S0168-9002(03)01368-8}{\textit{Nucl. Instrum. Meth. A} {\bf 506} (2003) 250} [\href{https://inspirehep.net/literature/593382}{\textsc{inSPIRE}}].

\bibitem{geant42} J. Allison {\it et al.}, \textit{Geant4 developments and applications}, \href{https://doi.org/10.1109/TNS.2006.869826}{\textit{IEEE Trans. Nucl. Sci.}  {\bf 53} (2006) 270} [\href{https://inspirehep.net/literature/715388}{\textsc{inSPIRE}}].

\bibitem{kkmc1} S. Jadach, B. F. L. Ward and Z. Was, \textit{The Precision Monte Carlo event generator {$\cal{KK}$} for two fermion final states in $e^+ e^-$ collisions}, \href{https://doi.org/10.1016/S0010-4655(00)00048-5}{\textit{Comput. Phys. Commun.}  {\bf 130} (2000) 260} [\href{https://arxiv.org/abs/hep-ph/9912214}{hep-ph/9912214}] [\href{https://inspirehep.net/literature/510990}{\textsc{inSPIRE}}].

\bibitem{kkmc2} S. Jadach, B. F. L. Ward and Z. Was, \textit{Coherent exclusive exponentiation for precision Monte Carlo calculations}, \href{https://doi.org/10.1103/PhysRevD.63.113009}{\textit{Phys. Rev. D} {\bf 63} (2001) 113009} [\href{https://arxiv.org/abs/hep-ph/0006359}{hep-ph/0006359}] [\href{https://inspirehep.net/literature/529540}{\textsc{inSPIRE}}].

\bibitem{evtgen1} R. G. Ping, \textit{Event generators at BESIII}, \href{https://doi.org/10.1088/1674-1137/32/8/001}{\textit{Chin. Phys. C} {\bf 32} (2008) 599} [\href{https://inspirehep.net/literature/1111724}{\textsc{inSPIRE}}].

\bibitem{evtgen2} D. J. Lange, \textit{The EvtGen particle decay simulation package}, \href{https://doi.org/10.1016/S0168-9002(01)00089-4}{\textit{Nucl. Instrum. Meth. A} {\bf 462} (2001) 152} [\href{https://inspirehep.net/literature/560129}{\textsc{inSPIRE}}].

\bibitem{PDG2022} \textsc{Particle Data Group} collaboration, \textit{Review of Particle Physics}, \href{https://doi.org/10.1093/ptep/ptac097}{\textit{PTEP} {\bf 2022} (2022) 083C01} [\href{https://inspirehep.net/literature/2106994}{\textsc{inSPIRE}}].

\bibitem{lundcharm} J. C. Chen {\it et al.} \textit{Event generator for $J/\psi$ and $\psi(2S)$ decay}, \href{https://doi.org/10.1103/PhysRevD.62.034003}{\textit{Phys. Rev. D} {\bf 62} (2000) 034003} [\href{https://inspirehep.net/literature/529201}{\textsc{inSPIRE}}].

\bibitem{photos2} E. Barberio, B. van Eijk and Z. Was, \textit{PHOTOS: A Universal Monte Carlo for QED radiative corrections in decays},
\href{https://doi.org/10.1016/0010-4655(91)90012-A}{\textit{Comput. Phys. Commun.} {\bf 66} (1991) 115} [\href{https://inspirehep.net/literature/299639}{\textsc{inSPIRE}}].

\bibitem{qed} E. A. Kuraev and V. S. Fadin, \textit{On Radiative Corrections to $e^+$ $e^-$ Single Photon Annihilation at High-Energy}, \href{}{\textit{Sov. J. Nucl. Phys.} \textbf{41} (1985) 466} [\href{https://inspirehep.net/literature/217313}{\textsc{inSPIRE}}].

\bibitem{isr} W. Sun {\it et al.}, \textit{An iterative weighting method to apply ISR correction to $e^+ e^-$ hadronic cross-section measurements}, \href{https://doi.org/10.1007/s11467-021-1085-6}{\textit{Front. Phys. (Beijing)} \textbf{16} (2021) 64501} [\href{https://arxiv.org/abs/2011.07889}{{arXiv:2011.07889}}] [\href{https://inspirehep.net/literature/1830422}{\textsc{inSPIRE}}].

\bibitem{vp} F. Jegerlehner and R. Szafron, \textit{$\rho^0$-$\gamma$ mixing in the neutral channel pion form factor $F^e_\pi(s)$ and its role in comparing $e^+ e^-$ with $\tau$ spectral functions}, \href{https://doi.org/10.1140/epjc/s10052-011-1632-3}{\textit{Eur. Phys. J. C} {\bf 71} (2011) 1632} [\href{https://arxiv.org/abs/1101.2872}{{arXiv:1101.2872}}] [\href{https://inspirehep.net/literature/884306}{\textsc{inSPIRE}}].

\bibitem{tracking1} BESIII collaboration, \textit{Evidence for $\psi'$ decays into $\gamma \pi^0$ and $\gamma \eta$}, \href{https://doi.org/10.1103/PhysRevLett.105.261801}{\textit{Phys. Rev. Lett.} {\bf 105} (2010) 261801} [\href{https://arxiv.org/abs/1011.0885}{{arXiv:1011.0885}}] [\href{https://inspirehep.net/literature/875391}{\textsc{inSPIRE}}].

\bibitem{tracking2} BESIII collaboration, \textit{Measurements of $\psi(3686) \to K^- \Lambda \bar{\Xi}^+$ and $\psi(3686) \to \gamma K^- \Lambda \bar{\Xi}^+$}, \href{https://doi.org/10.1103/PhysRevD.91.092006}{\textit{Phys. Rev. D} {\bf 91} (2015) 092006} [\href{https://arxiv.org/abs/1504.02025}{arXiv:1504:02025}] [\href{https://inspirehep.net/literature/1358401}{\textsc{inSPIRE}}].

\bibitem{pid1} BESIII collaboration, \textit{Observation of and improved measurement of $J/\psi \to p \bar{p} \eta'$}, \href{https://doi.org/10.1103/PhysRevD.99.032006}{\textit{Phys. Rev. D} {\bf 99} (2019) 032006} [\href{https://arxiv.org/abs/1812.05800}{arXiv:1812.05800}] [\href{https://inspirehep.net/literature/1709205}{\textsc{inSPIRE}}].

\bibitem{pid2} BESIII collaboration, \textit{Study of excited $\Xi$ states in $\psi(3686) \to K^-\Lambda\bar{ \Xi}^+~+~$ c.c.} \href{https://doi.org/10.1103/PhysRevD.109.072008}{\textit{Phys. Rev. D} {\bf109} (2024) 072008} [\href{https://arxiv.org/abs/2308.15206}{arXiv:2308.15206}] [\href{https://inspirehep.net/literature/2691955}{\textsc{inSPIRE}}].

\end{thebibliography}
\end{document}